\documentclass[onecolumn,amssymb,fleqn,12pt]{revtex4} %fleqn: make all eqns left alighed  ,showpacs
\usepackage{epsfig,amssymb,amsmath,graphicx,subfigure,hyperref}  % pdfpages and graphicx are added   hyperref(set hyperlink for content)
\usepackage{color}

\newcommand{\be}{\begin{eqnarray}}   %%\nonumber\\
\newcommand{\ee}{\end{eqnarray}}
\def\RR{\leavevmode\hbox{$\rm I\!R$}}
\def\I{{\cal{I}}}

\begin{document}

\title{Crystalline Particle Packings on Constant Mean Curvature (Delaunay) Surfaces}
\author{Enrique Bendito$^1$, Mark J. Bowick$^2$, Agustin Medina$^1$ and Zhenwei
Yao$^3$}
\affiliation{$^1$ Departament de Matem$\grave{a}$tica Aplicada III, Universitat
  Polit$\grave{e}$cnica de Catalunya, Spain\\
$^{2,}$$^3$Department of Physics, Syracuse University, Syracuse, New York
13244-1130,USA
}
\begin{abstract}
We investigate the structure of crystalline particle arrays on {\em constant mean curvature} (CMC) surfaces of revolution. Such curved crystals have been realized physically by creating charge-stabilized colloidal arrays on liquid  capillary bridges. CMC surfaces of revolution, classified by Charles Delaunay in 1841, include the 2-sphere, the cylinder, the vanishing mean curvature catenoid (a minimal surface) and the richer and less investigated unduloid and nodoid. We determine numerically candidate ground state configurations for 1000 point-like particles interacting with a pairwise-repulsive $1/r^3$  potential, with distance $r$ measured in 3-dimensional Euclidean space $\RR^3$. We mimic stretching of capillary bridges by determining the equilibrium configurations of particles arrayed on a sequence of Delaunay surfaces obtained by increasing or decreasing the height at constant volume starting from a given initial surface, either a fat cylinder or a square cylinder.
In this case the stretching process takes one through a complicated sequence of Delaunay surfaces each with different geometrical parameters including the aspect ratio, mean curvature and maximal Gaussian curvature.
Unduloids, catenoids and nodoids {\em all} appear in this process. 
Defect motifs in the ground state evolve from dislocations at the boundary to dislocations in the interior to pleats and scars in the interior and then 
isolated 7-fold disclinations in the interior as the capillary bridge narrows at the waist (equator) and the maximal (negative) Gaussian curvature grows. We also check theoretical predictions that the isolated disclinations are present in the ground state when the surface contains a geodesic disc with integrated Gaussian curvature exceeding $-\pi/3$. Finally we explore minimal energy configurations on sets of slices of a given Delaunay surface and obtain configurations and defect motifs consistent with those seen in stretching.     
\end{abstract}
\pacs{} 
\maketitle

\footnotetext[3]{Present address: Department of Materials Science and Engineering, Northwestern University, Evanston, IL 60208, USA}

\section{Introduction}

Much has been learned in the last decade about crystalline particle packings on
a wide variety of curved two-dimensional surfaces \cite{BG}.  The surfaces studied include the 2-sphere
(constant positive Gaussian curvature)~\cite{BNT, bowick2002crystalline, bausch2003grain,
bowick2007dynamics}, the 2-torus (variable positive and
negative Gaussian curvature with vanishing integrated Gaussian
curvature)~\cite{giomi2008defective, giomi2008elastic, pairam2009generation,yao2011shrinking}, the
paraboloid (variable Gaussian curvature and a
boundary)~\cite{giomi2007crystalline, bowick2008bubble}, the Gaussian bump 
(positive and negative Gaussian curvature)~\cite{sadoc1989infinite, vitelli2004defect,vitelli2006crystallography} and the catenoid minimal
surface~\cite{IVC, BY, irvine2012geometric}. 
Although the nature of condensed matter order on
surfaces is applicable to many different physical settings the richest
comparison between theoretical/numerical predictions and experiments has been
made with micron-scale colloidal systems.  In these systems solid colloidal particles
self-assemble at the interface of two distinct immiscible liquids, either in particle-stabilized (Pickering) emulsions~\cite{particlestabilizedemulsions,Pickering_1907,dinsmore2002colloidosomes}  or
charge-stabilized emulsions~\cite{LBHSC:2007,LZRCB:2007}. The
case of ordering on the 2-sphere (the surface of a ball) is realized by particles self-assembling
at the surface of a droplet held perfectly spherical by surface tension~\cite{bausch2003grain}. The
ordered configurations of particles may be imaged with confocal
microscopy and the particles even manipulated with laser
tweezers~\cite{IBCNatureMaterials}. In charge-stabilized emulsions particles do not even
wet the surface and so preserve the intrinsic shape of the droplet. The natural range in size $R$ of the droplets allows one to explore the
effect of the Gaussian curvature, which varies as $1/R^2$, on the ordering. The ground
state has been found to exhibit qualitatively different types of defect arrays as the size
changes~\cite{bowick2006crystalline, bausch2003grain}. Small droplets have
12 isolated 5-fold coordinated disclinations (5s) whereas for larger droplets the
isolated 5s sprout additional chains of dislocations (a dislocation is a tightly bound 
disclination-antidisclination or 5-7 pair) to become 12 grain boundary {\em scars}, each of which 
has a net disclination charge of $+1$ (i.e. one excess 5)  and freely 
terminates within the medium. Free termination of a grain boundary is possible 
in curved space because Bragg rows are curved along the converging geodesics of the 2-sphere, eventually healing the $30^{\circ}$ mismatch in crystallographic axes found at the center of the grain boundary. 
\begin{figure}[htb]
\begin{center}
\scalebox{.16}{{\includegraphics{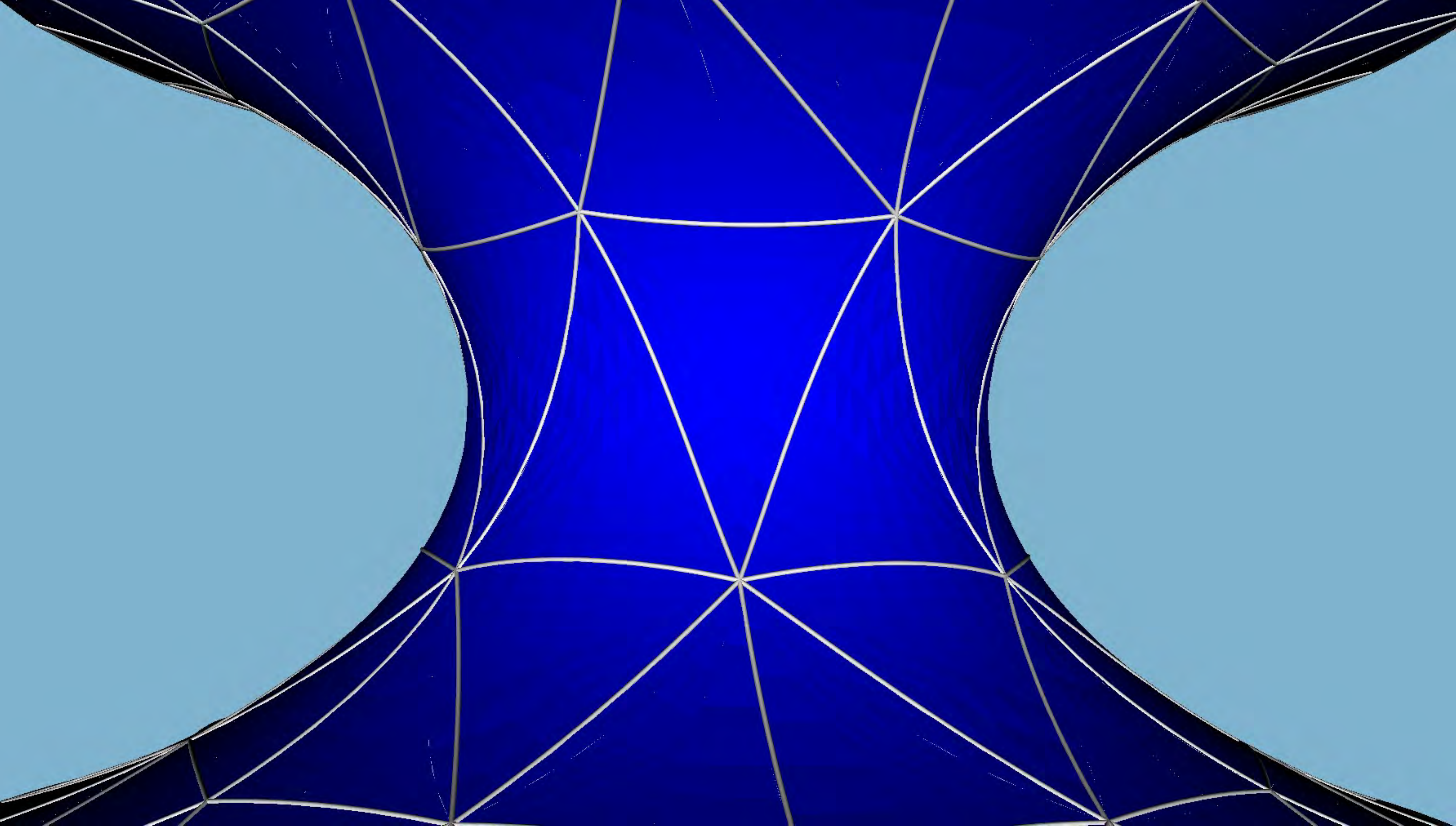}}}
\end{center}
\caption{\label{geodesics}  A geodesic triangulation of a negative Gaussian curvature nodoid: note the large departure of individual geodesic triangles from their Euclidean analogs where the interior angles add to $\pi$. }
\end{figure}

Recently a very rich and flexible experimental system has
been developed \cite{IVC} in which charge-stabilized emulsions are created in the form of
capillary bridges - these are structures in which a drop of liquid A, immersed in
liquid B, spans the gap between two parallel flat glass surfaces. The surface separating
the two liquids has the topology of a cylinder and necessarily has a {\em constant mean curvature} (CMC)
determined by the Laplace pressure difference between the inside (A) and
the outside (B) liquids. These CMC surfaces of revolution were classified in the 19th century by 
Charles Delaunay and come in 5 classes: the sphere, the cylinder, the catenoid, the unduloid 
and the nodoid~\cite{eells, Delaunay, Sturm}. Negative Gaussian curvature is expected to give rise to 
quite different structures from surfaces with positive Gaussian curvature
\cite{nelson1983order,RN,BNT,TSV,STN,IVC,irvine2012geometric,BY,KW}. The simplest negative curvature surface 
is the constant negative curvature hyperbolic plane $H^2$, the negative curvature analog of the 2-sphere $S^2$,
but the isometric embedding of $H^2$ as a complete subset of Euclidean 3-space is not differentiable \cite{HW} 
and may not be realizable physically. To illustrate one elementary feature of negative curvature surfaces  we
highlight in Fig.~\ref{geodesics} a single geodesic triangle (in red) on a negative
Gaussian curvature nodoid (blue)-- the sum of the interior angles of this triangle is
$139^{\circ}$, which is a striking departure from the $180^{\circ}$ sum of a Euclidean triangle.
Note also the diverging geodesics on the surface. The 

In this paper we explore the structure of the crystalline ground state of
particles strictly confined to a Delaunay CMC surface and interacting with a
pairwise-repulsive short-range power law potential. Such surfaces have varying negative Gaussian 
curvature and are thus technically more challenging to analyze and simulate numerically. 
We are particularly interested in the defect structure of the ground state and how distinctive
defect motifs emerge as the integrated Gaussian curvature is varied within one
class of CMC surface as well as the evolution of the ground state as one quasi-statically 
varies the manifold by increasing the height of the capillary bridge: the
sequence of Delaunay surfaces studied corresponds to the 
physical experiment of pulling the bounding plates slowly apart or pushing them slowly together
and hence stretching or compressing the capillary bridge.

\section{Geometry of Delaunay surfaces}

Delaunay surfaces are variationally determined by being constant mean curvature surfaces of revolution
with minimum lateral area and fixed volume. The characterization
that most directly gives rise to a simple and useful parametrization, however, 
is to consider them as the surfaces of revolution whose meridians are the roulettes 
of the foci of the conics, that is the curves traced by the foci of the
conics as they roll on a conic tangent without slip (for a detailed study
see~\cite{B13}). Besides the extreme 
limiting cases of the sphere and the cylinder these surfaces are catenoids,
unduloids and nodoids,  generated by the roulettes of
parabolas, ellipses and hyperbolas respectively.

\begin{figure}[htb]
\begin{center}
\scalebox{.15}{{\includegraphics{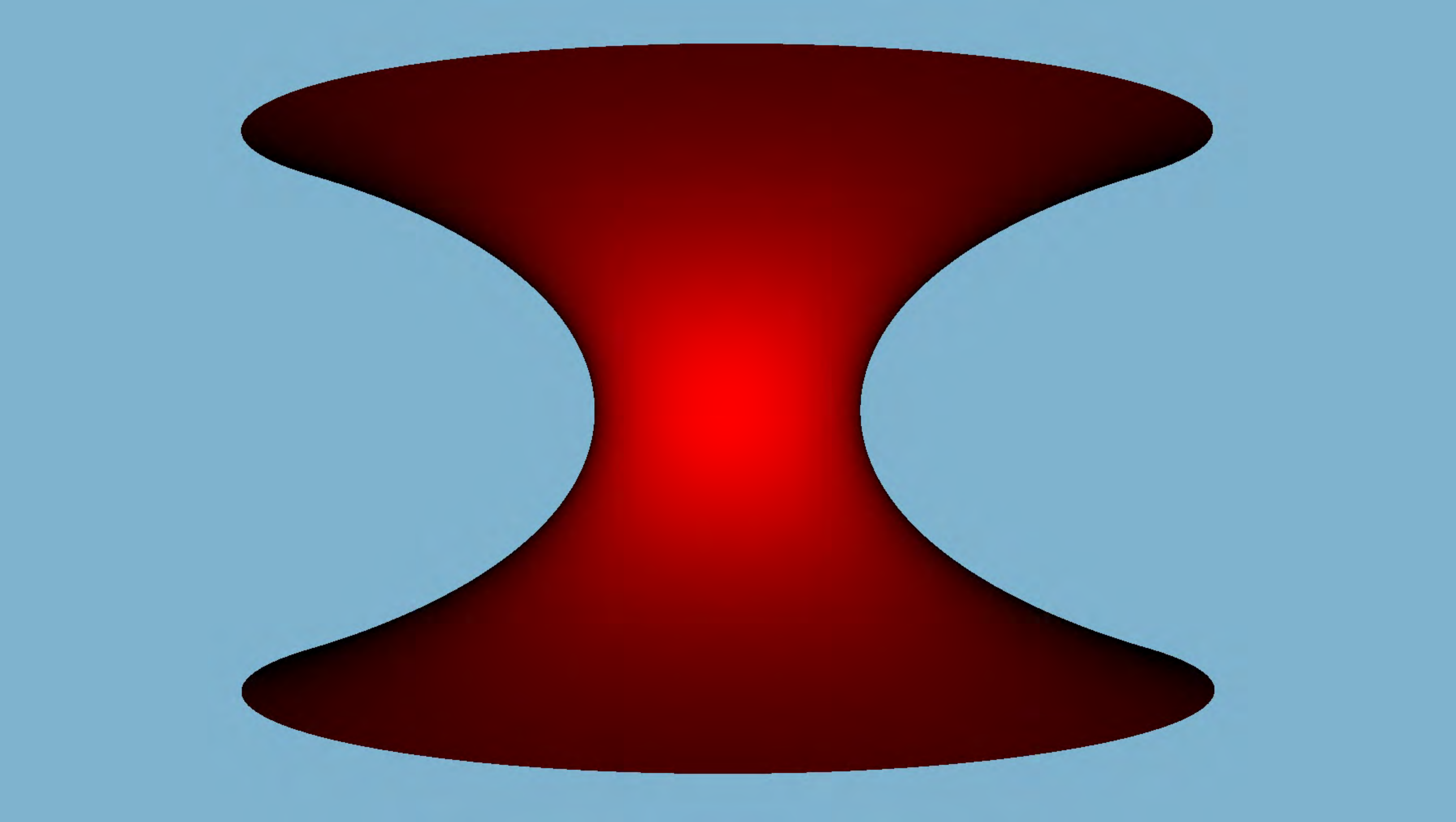}}}
\end{center}
\caption{\label{catenoid} A catenoid.}
\end{figure}

The roulette of the focus of the parabola is the catenary, and the 
surface of revolution, the catenoid, is the best known nontrivial Delaunay
surface. Its parameterization is (see Fig~ \ref{catenoid}):
\begin{equation}
\centering
\label{cate}
\pmb x(t,v)=\displaystyle\left(c\cosh\left(\frac{t}{c}\right)\cos(v),
c\cosh\left(\frac{t}{c}\right)\sin(v),t\right),
\end{equation} 
where $c$ is the radius of the waist, $v\in (0,2\pi)$ and $t\in\RR$.

The parameterization of an unduloid is given by
\begin{equation}\label{undu}
\begin{array}{rl}
\pmb y(t,v)&=\left( f_u(t)\cos(v), f_u(t)\sin(v),g_u(t) \right) \ ,\\[2ex]
\hbox{where}\,\,\,f_u(t)&=\displaystyle\frac{b(a-c\cos(t))}{\sqrt{a^2-c^2\cos^2(t)}},\\[3ex]
g_u(t)&=\displaystyle\int_{t_0}^t\sqrt{a^2-c^2\cos^2(z)}\,dz-\frac{c\sin(t)\left(a-c\cos(t)\right)}
{\sqrt{a^2-c^2\cos^2(t)}}, \\
\end{array}
\end{equation}
%\eqref{undu} 
$a$ and $b$ are the semiaxes of the ellipse, $c=\sqrt{a^2-b^2}$ and
$v\in(0,2\pi)$.  If $t\in \left(-\frac{\pi}{2},\frac{\pi}{2}\right)$, the
roulette generated by the closest focus to the tangent to the ellipse generates
unduloids of the type shown in Fig.~\ref{undu} (left) 
with negative Gaussian curvature. Taking, instead, $t\in \left(\frac{\pi}{2},\frac{3\pi}{2}\right)$, 
the same focus generates the unduloids with positive Gaussian curvature, such as the one 
shown in Fig.~\ref{undu} (right). 
The same surfaces are obtained by using the other focus, but this time in the 
reverse order.
\begin{figure}[htb]
\begin{center}
\scalebox{.12}{{\includegraphics{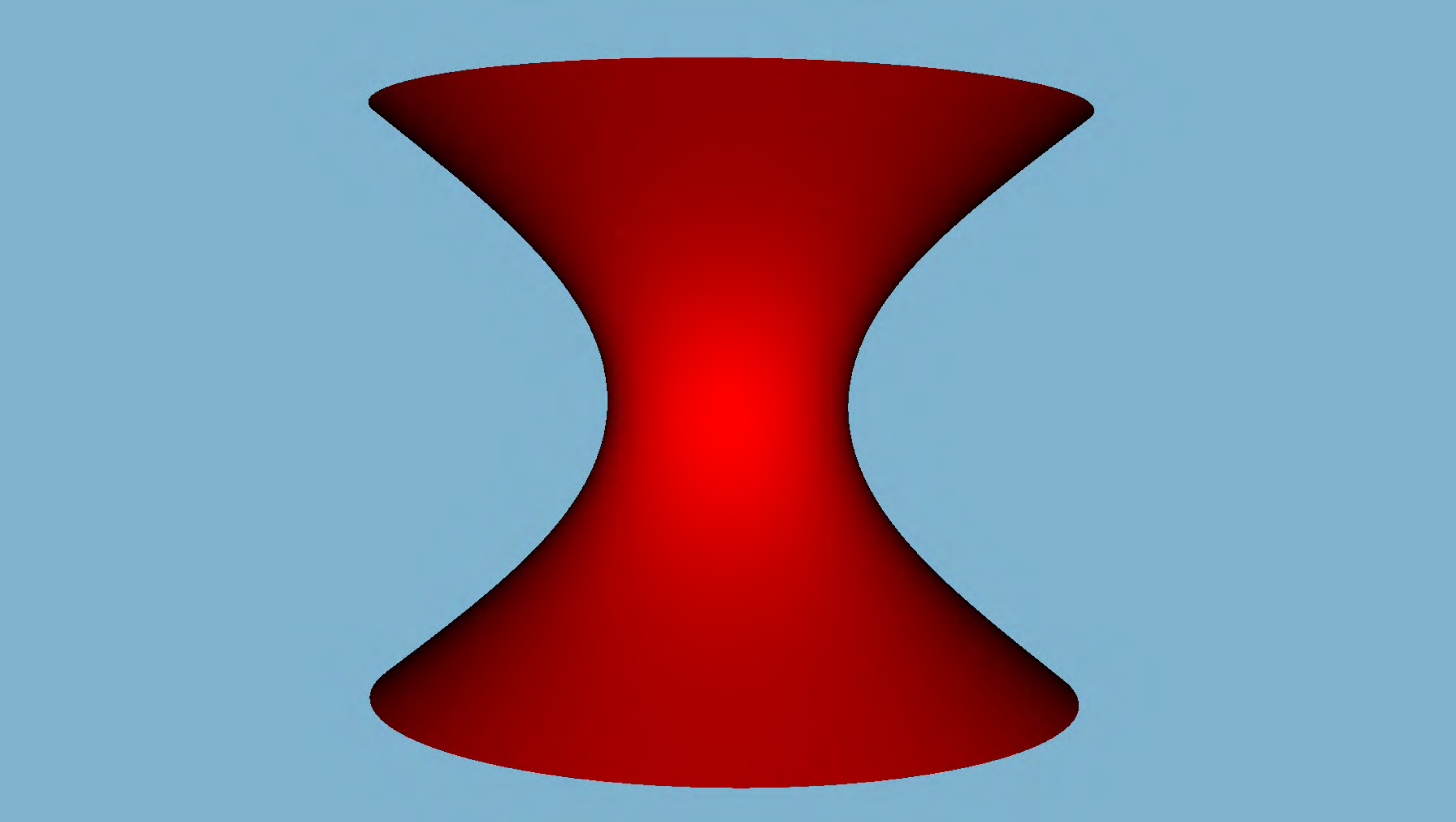}}}
\scalebox{.12}{{\includegraphics{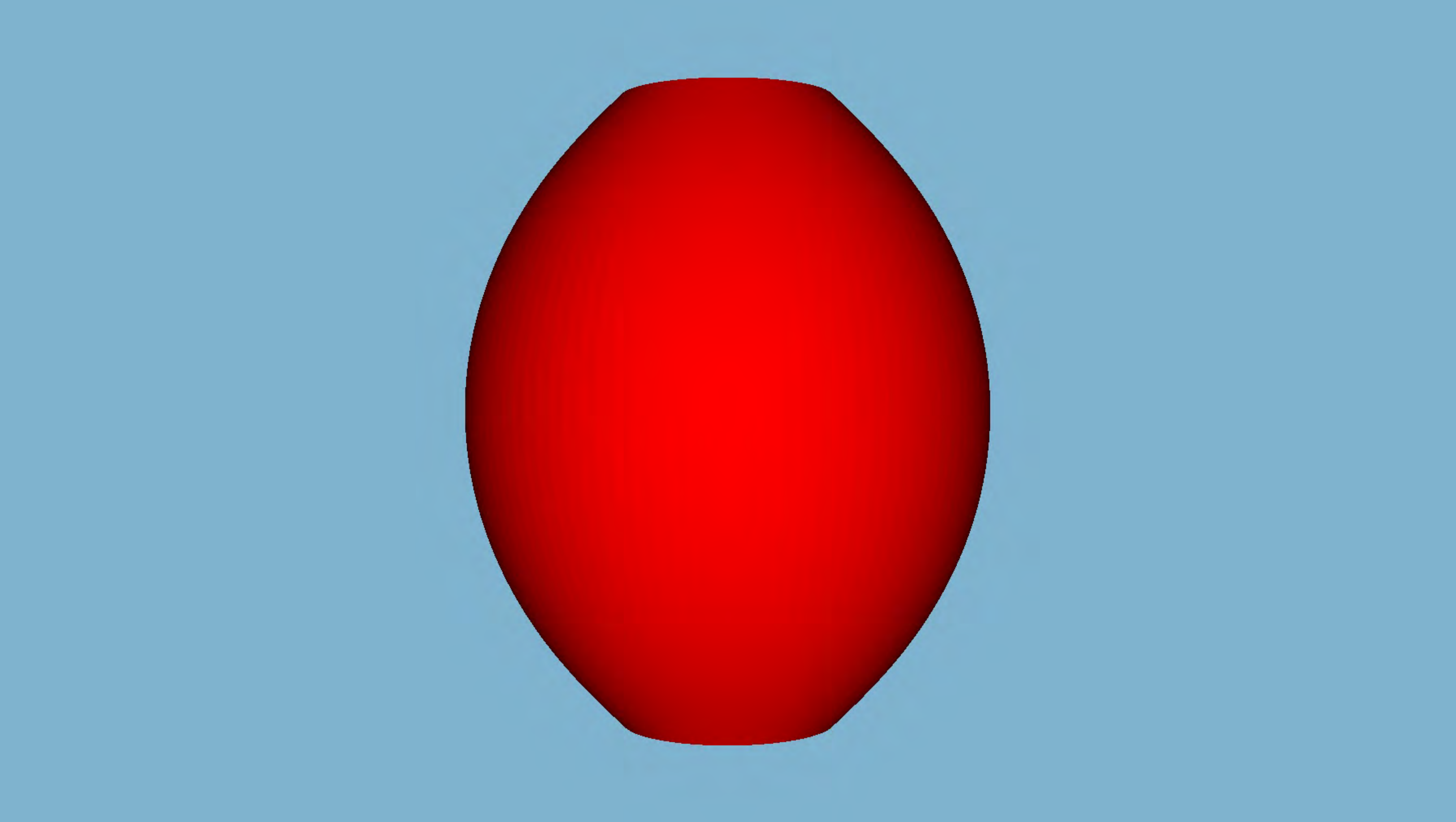}}}
\end{center}
\caption{\label{undu} Unduloid with negative Gaussian curvature (left) and positive Gaussian curvature (right).}
\end{figure}

\begin{figure}[htb]
\begin{center}
\scalebox{.12}{{\includegraphics{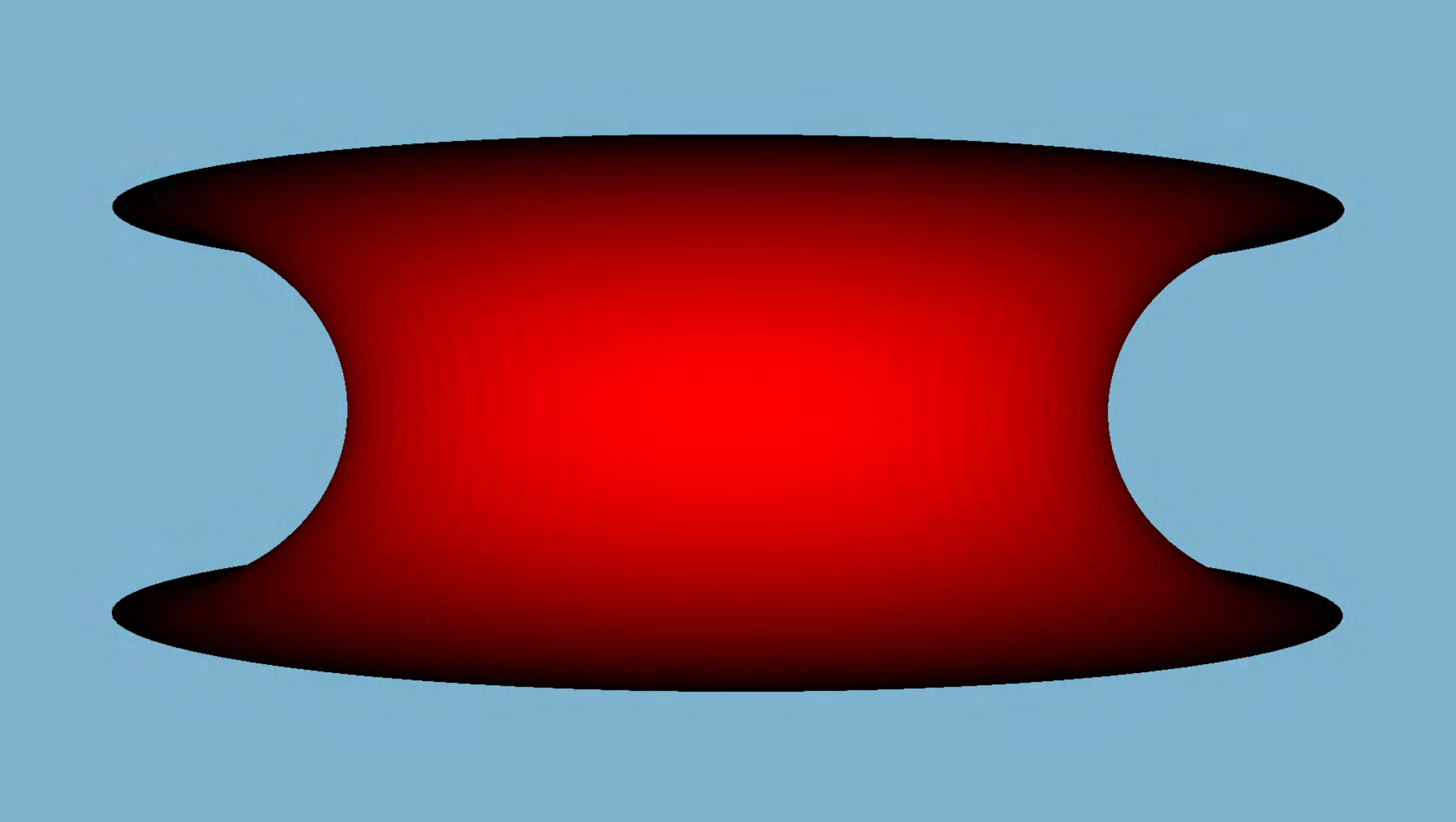}}}
\scalebox{.12}{{\includegraphics{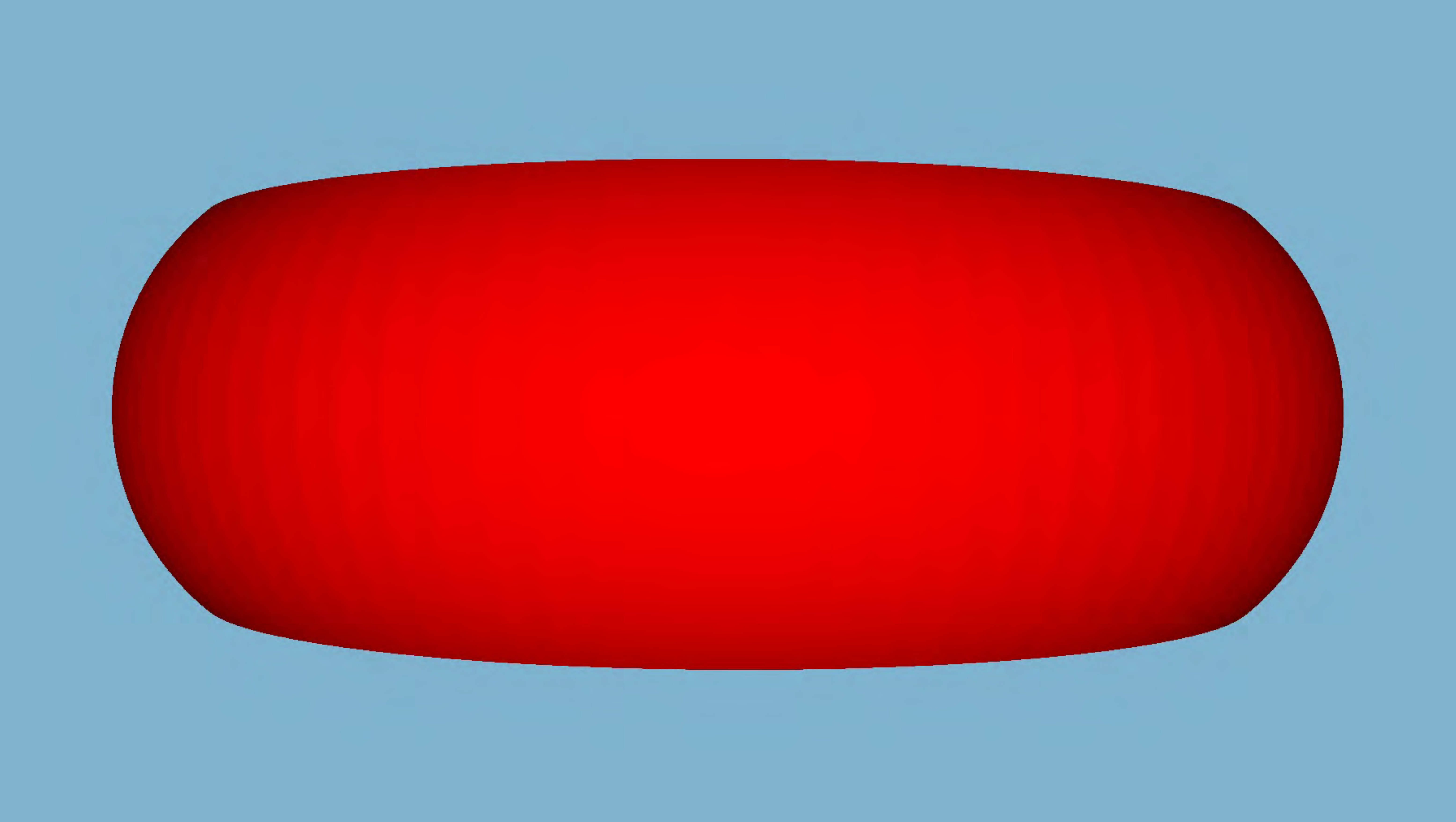}}}
\end{center}
\caption{\label{nodo} Nodoid with negative Gaussian curvature (left) and positive Gaussian curvature (right).}
\end{figure}

Similarly, the parameterization of a nodoid is given by
\begin{equation}\label{nodo1}
\begin{array}{rl}
\pmb z^1(t,v)&=\left( f^1_n(t)\cos(v), f^1_n(t)\sin(v),g^1_n(t) \right) \ ,\\[2ex]
\hbox{where}\,\,\,f^1_n(t)&=\displaystyle\frac{b(c\cosh(t)-a)}{\sqrt{c^2\cosh^2(t)-a^2}},\\[4ex]
g^1_u(t)&=\displaystyle\int_{t_0}^t\sqrt{c^2\cosh^2(z)-a^2}\, dz-\frac{c\sinh(t)\left(c\cosh(t)-a\right)}
{\sqrt{c^2\cosh^2(t)-a^2}}, \\
\end{array}
\end{equation}
where $a$ and $b$ are the semiaxes of the hyperbola, 
$c=\sqrt{a^2+b^2}$, $v\in(0,2\pi)$ and $t\in \RR$. 
The expressions in $(3)$ treat the roulette generated 
by the closest focus to the tangent to the hyperbola and 
yield nodoids with negative Gaussian curvature, as shown in 
Fig.~\ref{nodo} (left). Taking the roulette generated by the 
other focus yields the expressions $(4)$ and nodoids with positive 
Gaussian curvature, as illustrated in Fig.~\ref{nodo} (right).
\begin{equation}\label{nodo2}
\begin{array}{rl}
\pmb z^2(t,v)&=\left( f^2_n(t)\cos(v), f^2_n(t)\sin(v),g^2_n(t) \right)\\[2ex]
\hbox{where}\,\,\,f^2_n(t)&=\displaystyle\frac{b(c\cosh(t)+a)}{\sqrt{c^2\cosh^2(t)-a^2}},\\[4ex]
g^2_u(t)&=\displaystyle\int_{t_0}^t\sqrt{c^2\cosh^2(z)-a^2}\, dz-\frac{c\sinh(t)\left(c\cosh(t)+a\right)}
{\sqrt{c^2\cosh^2(t)-a^2}}. \\
\end{array}
\end{equation}

All the fundamental properties of the Delaunay surfaces can be 
obtained easily using these parameterizations. In particular, the mean curvature 
of catenoids, unduloids and nodoids is given by
\begin{equation}\label{cm}
H_c=0,\hspace{1cm} H_u=\frac{1}{2a},\hspace{1cm} H^1_n=H^2_n=\frac{-1}{2a},
\end{equation}
respectively, where $a$ is the semi-major axis of the related conic. The Gaussian curvatures of catenoids, unduloids and nodoids are given by
\begin{equation}\label{cg}
\begin{array}{rllr}
K_c=&\displaystyle\frac{-1}{c^2 \cosh^4(t)}  \hspace{.5cm} &K_u=\displaystyle \frac{-c\cos(t)}{a\left(a-c\cos(t)\right)^2}\\[3ex]
K^1_n=&\displaystyle\frac{-c\cosh(t)}{a\left(c\cosh(t)-a\right)^2} \hspace{.5cm}
&K^2_n=\displaystyle \frac{c\cosh(t)}{a\left(c\cosh(t)+a\right)^2}\, .\\   
\end{array}
\end{equation}

\section{Numerical simulation: Finding a minimum energy configuration on a
Delaunay surface}

To determine candidate minimum energy configurations we use the Forces Method \cite{B07,B08}.
The basic structure of the Forces Method is classical
and explained in detail in \cite{B08}. For completeness we give a brief 
description here. It can be viewed as a local relaxation-gradient-like descent algorithm 
where each step consists of finding the update direction and the step size in a
predefined way. It consists of four steps:
\begin{itemize}\parskip=-.1cm
\item[$\bullet$] Choose a certain number of particles and an initial configuration for them
\item[$\bullet$] Update the positions of the particles in three-dimensional space
\item[$\bullet$] Project the positions on the surface
\item[$\bullet$] Repeat steps 2 and 3 until a given threshold is reached
\end{itemize}

\noindent Briefly these are:

\noindent
{\bf Initial configuration.} The initial configuration is chosen to be as uniformly distributed on the given Delaunay surface as possible, with the restriction that the particles are not near the two boundaries.  
Many independent runs (order 10) are made starting from different initial configurations and the lowest energy  configuration is selected.

\noindent
{\bf Update and projection.} For one particle
$$\hat{{\vec x}}^{k+1}={\vec x}^k+\lambda_k \, {\vec w}^k,$$ 
where ${\vec x}^k$, ${\vec w}^k$ and $\lambda_k$ are the position, update direction and 
step size at the $k$th step, respectively. $\hat{{\vec x}}^{k+1}$ would be
the new location of the particle if this location were on the surface.  Since it is generically off the surface after the update 
the actual position ${\vec x}^{k+1}$ is obtained by projecting
$\hat{{\vec x}}^{k+1}$ onto the surface.

The update direction is in the direction of the net force on the particle following from the gradient of the potential. In this paper, as noted before, we use a $1/r^3$ power law potential. For 
a system of $N$ charge 1 particles, $\vec x_i\in \RR^3,$
$i=1,\ldots,N,$ the potential energy is then
given by $\I_N=\displaystyle\frac{1}{2}\sum\limits_{i=1}^{N}V_i$, where
$V_i=\displaystyle\sum\limits_{j=1 \atop j\not=i}^{N} 
\frac{1}{|\vec x_i-\vec x_j|^3}$ 
is the potential created at $\vec x_i$ by all the other particles. We denote by
$\vec F_i$ minus the gradient of $V_i$ in terms of the position of the $i$-th
particle. 

If the particles lie on a regular surface $S$, then equilibrium is reached when the component of $\vec F_i$ tangential to $S$, $\vec F_i^T$, vanishes at $\vec x_i$. Then we choose $w=(\vec w_1,\ldots,\vec w_N)$ as the step direction, where
$\vec w_i=\displaystyle\frac{\vec F_i^T}{|F_{max}|}$ (we normalize by the maximum force).
The magnitude of the step size is then
$\lambda_k(\vec x_i)=d\left(\min\limits_{1\le j\le N\atop j\not=i}\{|\vec x_i^k-\vec x_j^k|\}\right)^2$, where the coefficient $d$ is a constant positive scalar. 
The existence of a minimum distance between particles makes it possible to adjust the step size to most efficiently access 
the various configurations that arise during the iterative process. The error at iteration $k$ is $w_{\rm max}=\max\limits_{1\le i\le N}{|\vec w_i|}$.
The algorithm stops when $w_{\rm max}$ reaches a certain prescribed threshold value $\varepsilon>0$. 

The 1000 particle simulations are performed on an Intel core i7 processor at a
speed of 36.5 iterations/sec. We simulate 23 distinct particle systems in all
requiring a total of $5.522\times 10^6$ iterations for a total CPU time of 42
hours$^1$. \footnotetext[1]{Coordinates and energies for the configurations analyzed in this paper may be obtained by contacting Mark Bowick (bowick@phy.syr.edu).}

\section{Constant volume results}

In this section we analyze the evolution of defect motifs on a capillary bridge
as it is stretched while preserving the volume and the contact area with both bounding
parallel flat plates (corresponding to the boundary disks of the Delaunay surface). 
At each stage of stretching the capillary bridge is a distinct Delaunay surface
with different mean and Gaussian curvatures. Thus, in the experimental setting,
the particles must continuously re-equilibrate to a new ground state.  Numerically we minimize the energy on each Delaunay surface separately. This analysis allows one to explore the structure of an entire sequence of Delaunay surfaces, one for each 
stage of the stretch. We study two types of constant volume stretching: one starts
from a fat cylindrical capillary bridge and the other from a square cylindrical capillary bridge.

\subsection{Stretching from an initial fat cylindrical capillary bridge}

\begin{figure}[htb]
\begin{center}
\scalebox{.6}{{\includegraphics{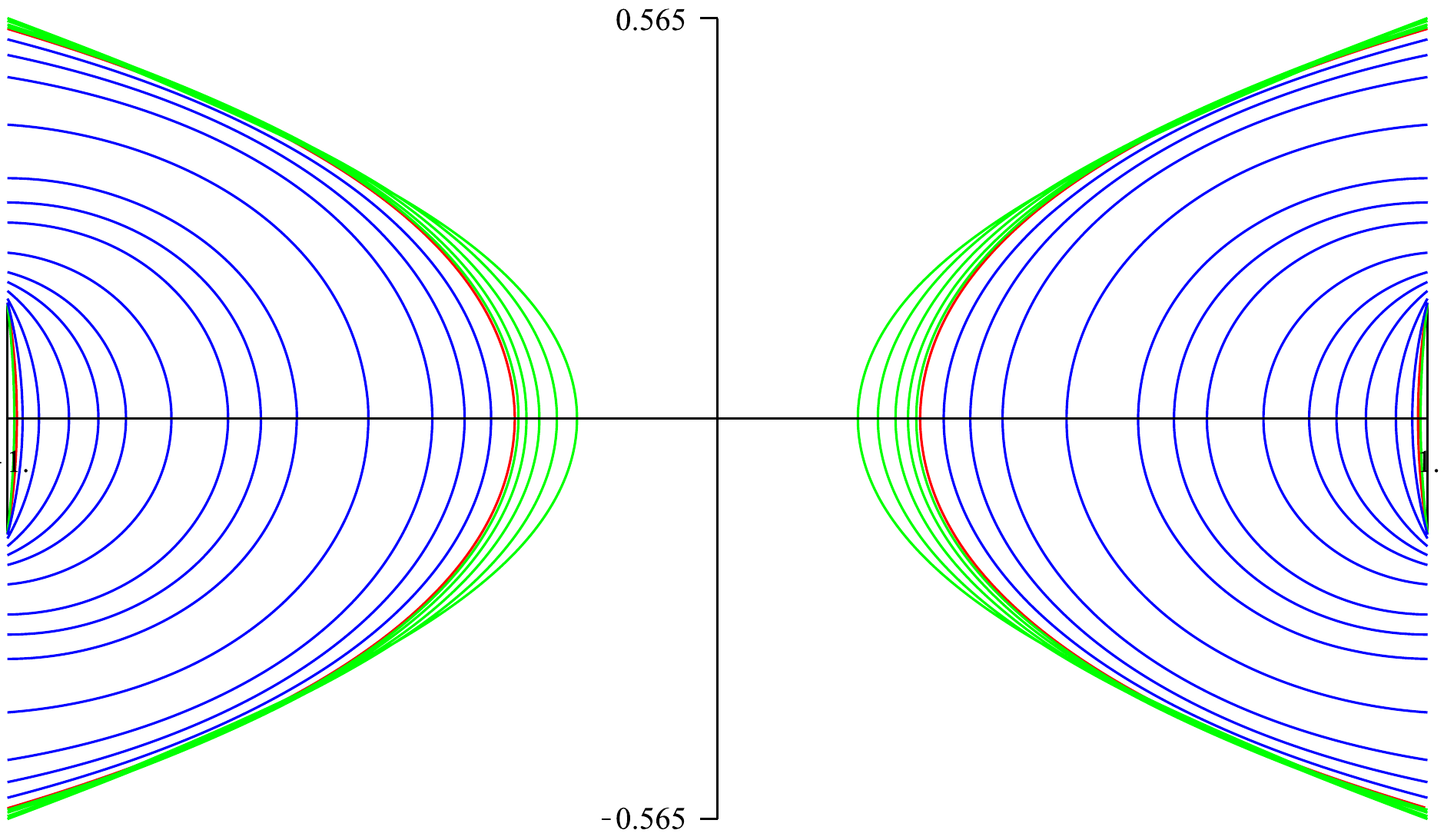}}}
\end{center}
\caption{\label{family} Representative members of the family of surfaces obtained by stretching a fat cylinder. From the outside-in: cylinder (black), unduloids (green), catenoid (red), nodoids (blue), catenoid (red) and unduloids (green).}
\end{figure}

The initial fat cylinder has contact radius $r_c=1$ and height
$h=1/\pi$, so both the volume and the aspect ratio (the ratio $\rho$ of the
radius of the surface at an equatorial plane (waist) to the radius of the contact disk
at the plates) are unity. In our numerical study starting from a fat cylindrical capillary bridge  
yields a rich family of surfaces that begins with the cylinder and subsequently deforms to a series of 
unduloids, the catenoid and then a series of nodoids. 
Thus all the classes of Delaunay surfaces, apart from the well-studied 2-sphere
which can be generated by other simpler means, are explored in this single process! Minimum energy configurations for charged particles interacting via a Yukawa potential and confined to Delaunay surfaces were obtained in \cite{KW}, but only for catenoids and unduloids. As is clear from Fig.~\ref{family}, the predominant surfaces in stretching are nodoids.
The proper treatment of nodoids is therefore essential  to a comparison with the experiments of Ref.\cite{IVC}.
Planar sections that contain the common revolution axis of several surfaces 
from this family are shown in Fig.~\ref{family}. 
The final unduloid has an aspect ratio $\rho = 0.198$.

We obtain minimum energy states for $N=1000$ particles interacting via a 
pairwise-repulsive $1/r^3$ potential. Short-range potentials (decaying faster than $1/r^2$) 
produce more uniform ground state configurations because long-range potentials 
(decaying slower than $1/r^2$) drive particles to the boundary circles to
maximize the average separation. This can lead to defect structures that are
influenced by the boundary {--} while of intrinsic interest as a boundary driven
phenomenon this is not our primary focus in this paper. We found that 
long-range effects may also be reduced, but not eliminated, by adding a line of neutralizing charge along the central axis of the surface. Here we focus on the role of variable Gaussian curvature  
in the interior of Delaunay surfaces. 
 
Various defect motifs emerge as the capillary bridge becomes higher and thinner (decreasing aspect ratio with our definition). Basically we identify the following sequence: at $\rho=0.984$, individual dislocations (tightly bound 5-7 pairs) appear at the boundary, resulting from the repulsion between dislocations of identical orientation. 

\begin{figure}[htb]
\begin{center}
\scalebox{.12}{{\includegraphics{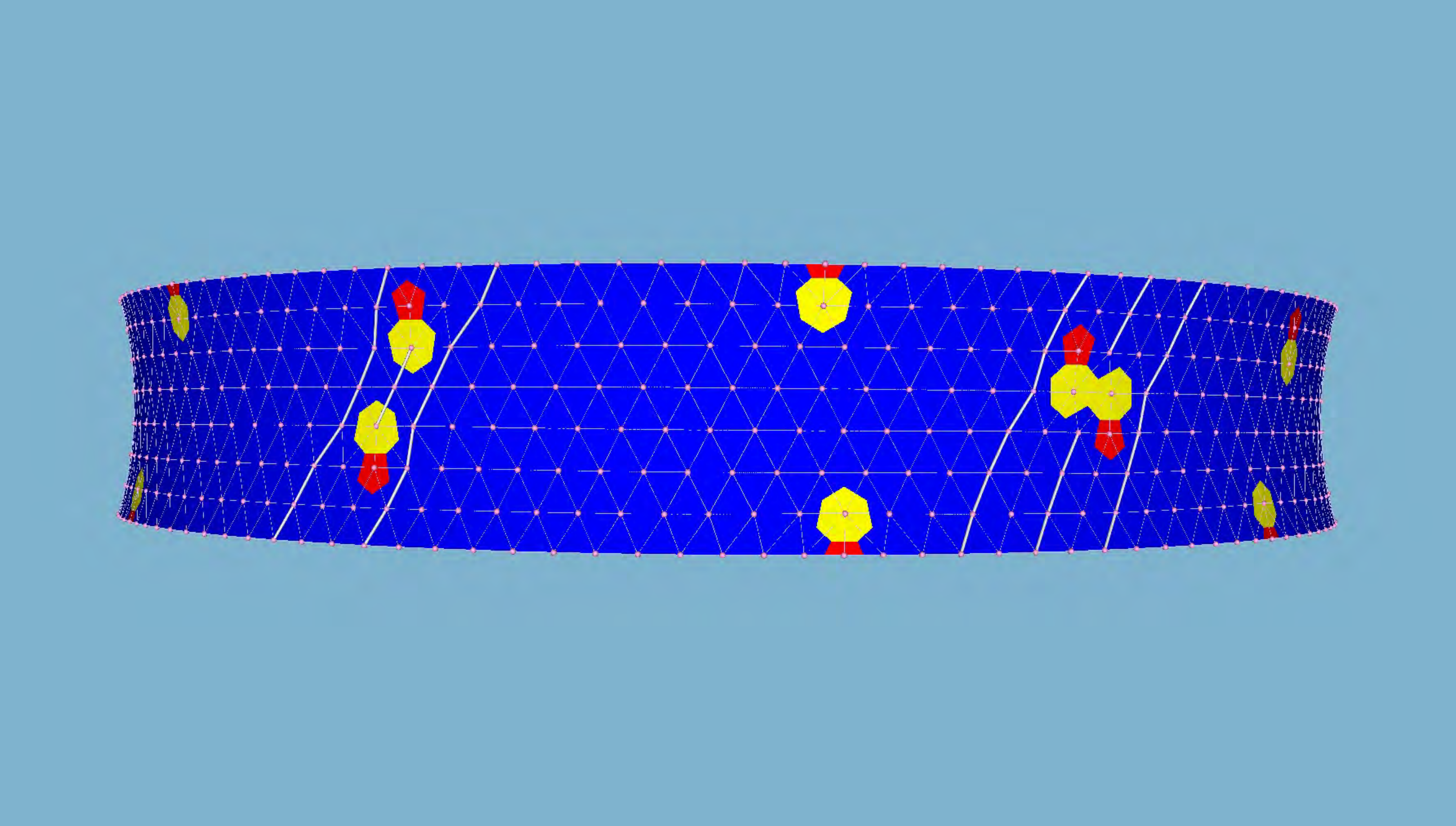}}}
\scalebox{.12}{{\includegraphics{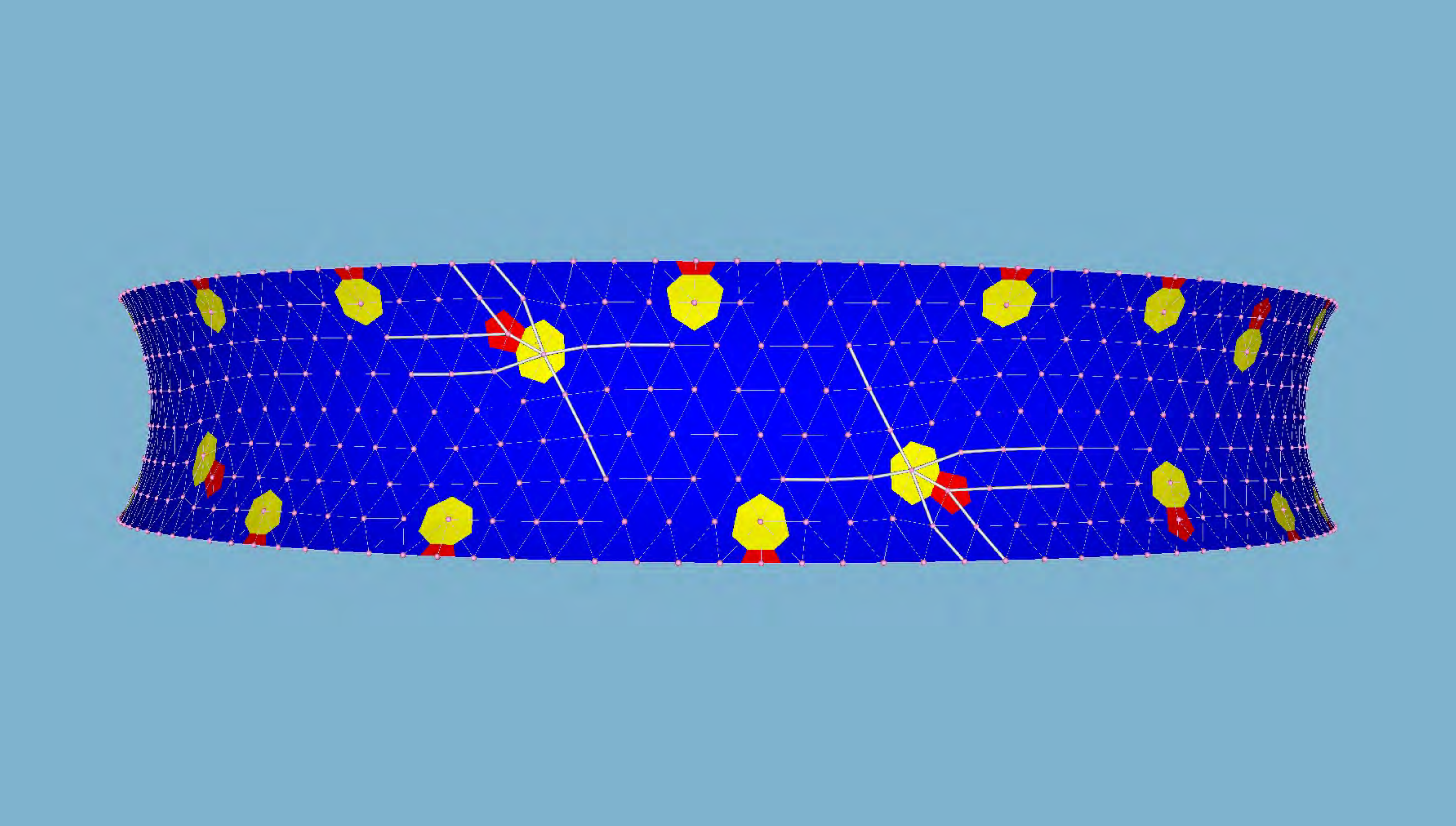}}}
\end{center}
\caption{\label{95_97} Left: Equilibrium configuration for a nodoid with aspect ratio $\rho=0.97$. Some Bragg rows are highlighted in white to illustrate the insertion of half-lines of particles defining a dislocation.  
Right: Equilibrium configuration for a nodoid with $\rho=0.95$.  Some dislocations are parallel to the nodoid's boundary.}
\end{figure}

\begin{figure}[htb]
\begin{center}
\scalebox{.12}{{\includegraphics{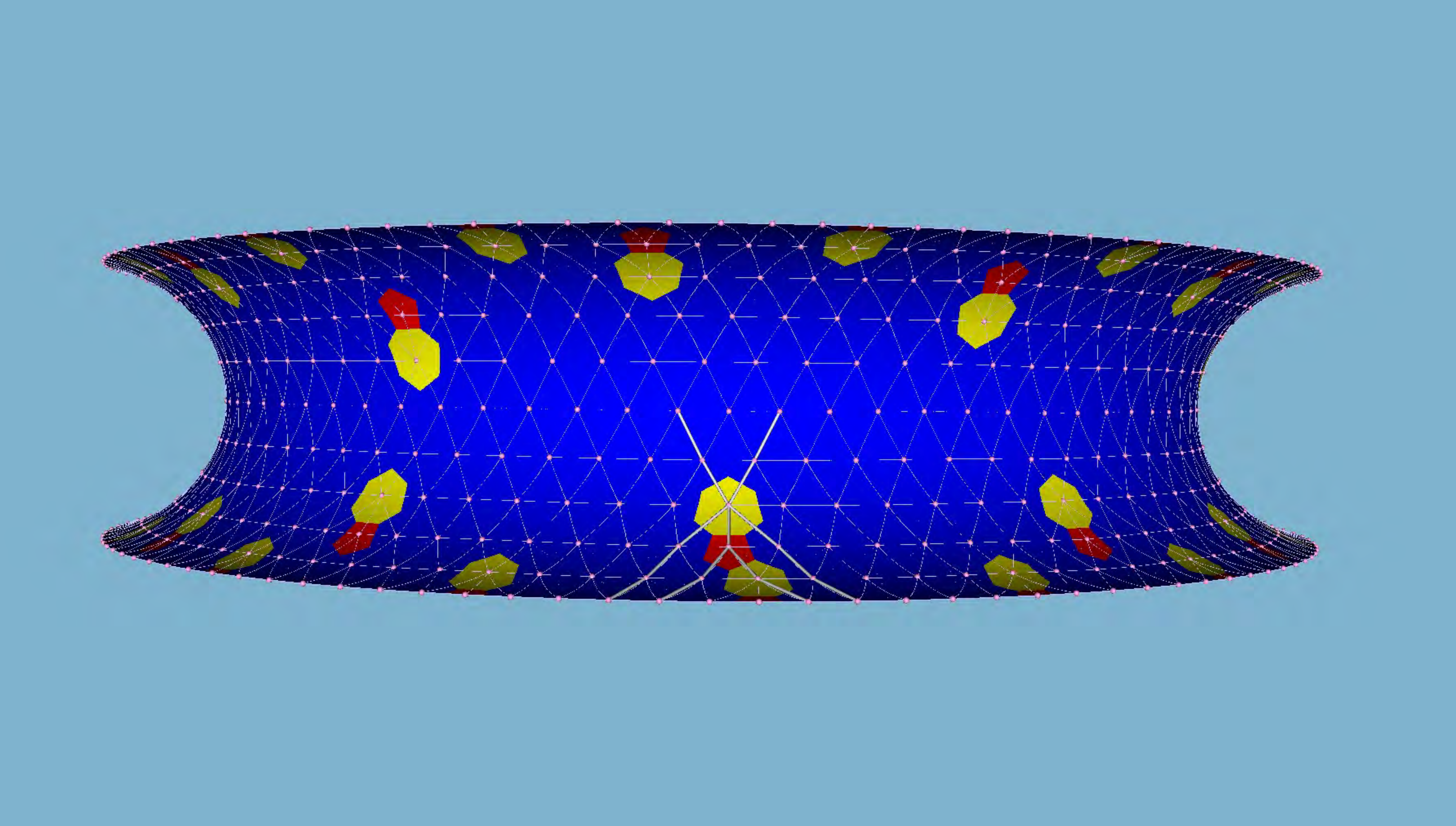}}}
\scalebox{.12}{{\includegraphics{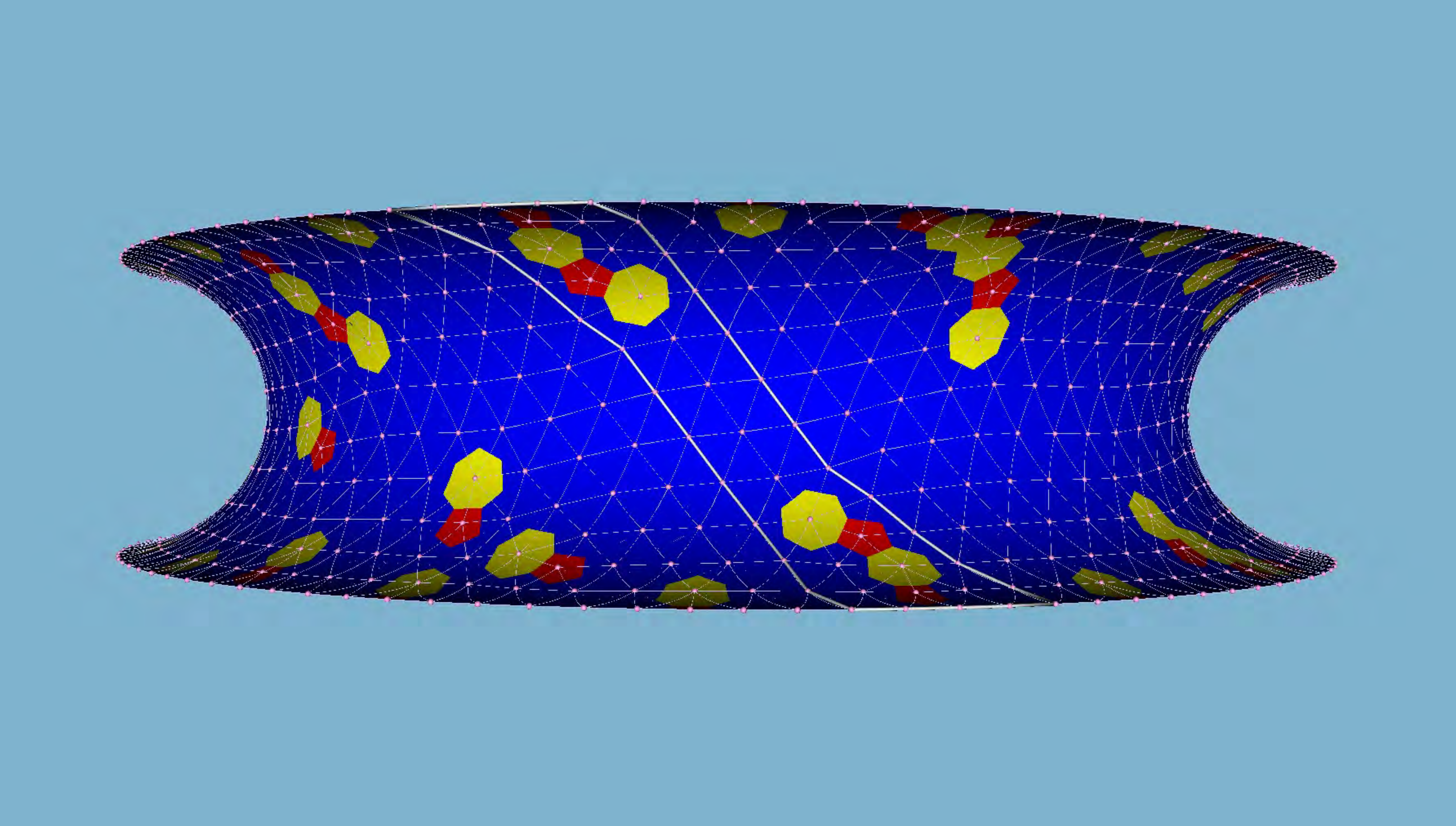}}}
\end{center}
\caption{\label{lin81_78} Left: Appearance of pleats at aspect ratio $\rho=0.81$. The distortion of the crystallographic rows corresponds to that of a chain of dislocations. Right: At 
$\rho=0.78$ pleats and dislocations begin to proliferate.}
\end{figure}

As $\rho$ decreases, these dislocations migrate to the interior of the surface and occasionally form multiple dislocation clusters. 
In Fig.~\ref{95_97}(left, right) we show minimum energy configurations for $\rho=0.97$ and $\rho=0.95$. The dual pentagons and heptagons shown here are obtained by first performing a Delaunay
triangulation of the configuration and then connecting the barycenters of adjacent triangles.  In Fig.~\ref{95_97}(left) we also highlight the Bragg rows each side of two pairs of dislocations so that one 
can see the removal (insertion) of a half-line of particles characteristic of a dislocation (anti-dislocation). 
In Fig.~\ref{95_97}(right) we highlight the Bragg rows surrounding two dislocations oriented almost tangentially to the boundary (and thus with Burgers vector almost parallel to the boundary).  
At $\rho=0.81$, pleats appear attached to the boundary, as shown in
Fig.~\ref{lin81_78}(left). Pleats are neutral grain boundaries with a 5-fold disclination at one end and a 7-fold disclination at the other end. A pleat can freely terminate at one or both ends and so they are easy to recognize.
Below $\rho=0.78$ pleats and dislocations proliferate both in the interior and at the boundaries (see
Fig.~\ref{lin81_78}(right)). 
 
At $\rho=0.774$ one sees for the first time isolated 7-fold disclinations (7s). In the figures these are seen clearly in the form of their dual heptagons. This is the true hallmark of the prevailing negative Gaussian curvature in the interior of the surface! 7-fold disclinations may arise by the unbinding  of a 7 from a pleat, which typically end with a 7 oriented towards the central interior of the surface. Indeed there are usually compensating scars with a net positive disinclination charge near an isolated 7 in this regime of aspect ratios. As the negative Gaussian curvature increases, however, isolated 7s are typically farther away from the positive scars that compensate them topologically.
\begin{figure}[htb]
\begin{center}
\scalebox{.12}{{\includegraphics{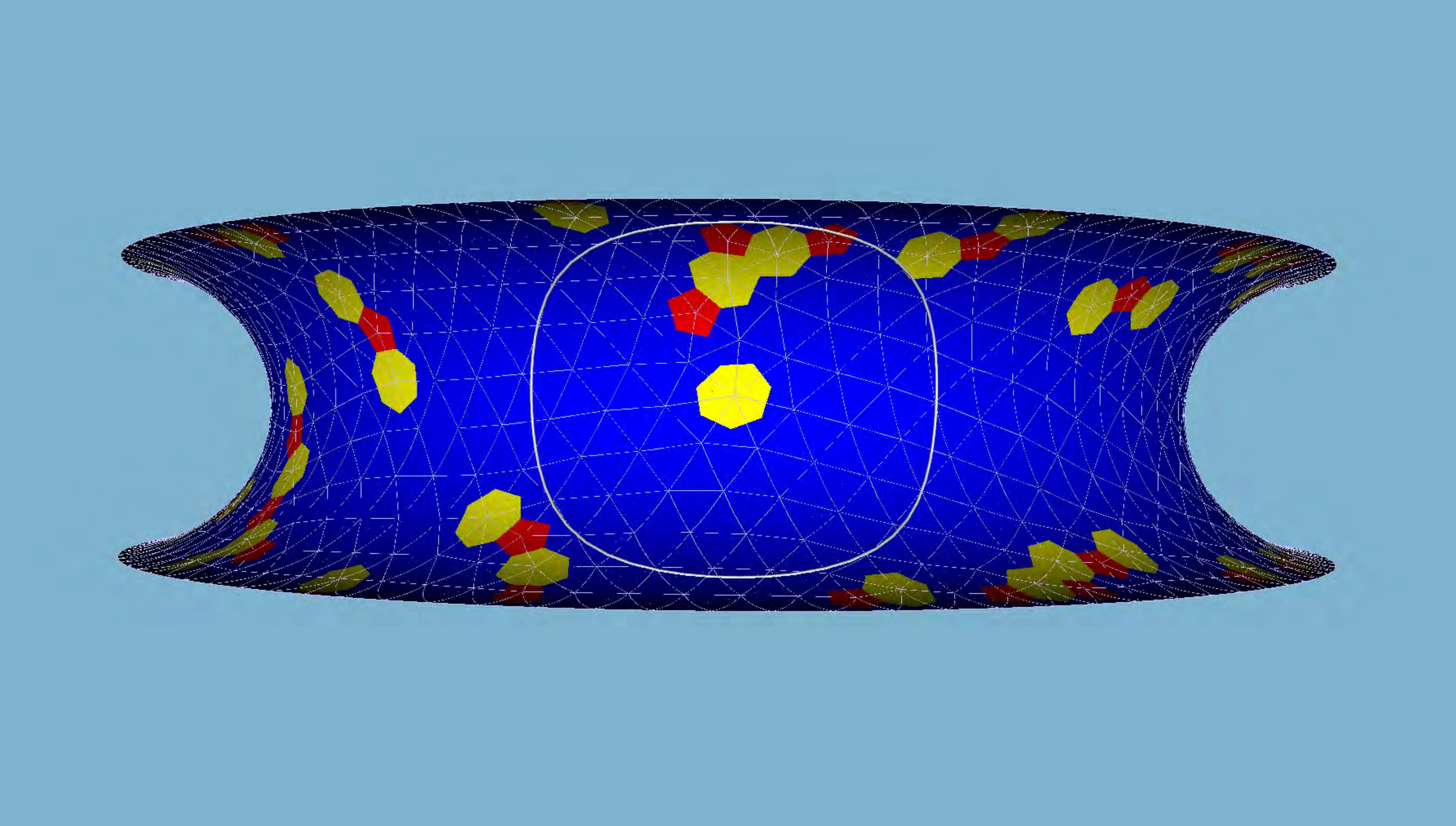}}}
\scalebox{.12}{{\includegraphics{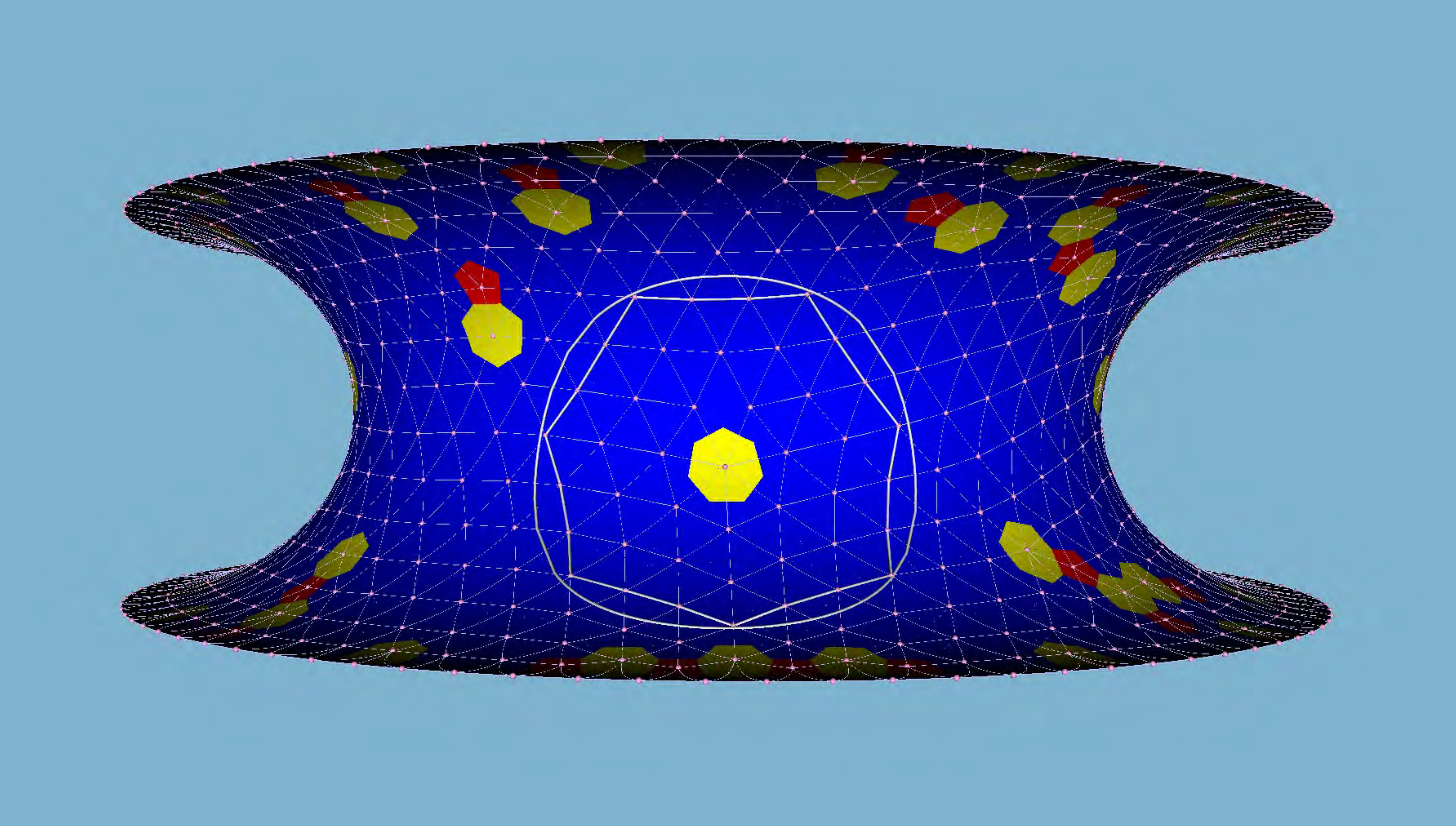}}}
\end{center}
\caption{\label{cir} Left: Nodoid with aspect ratio $\rho=0.774$. Note the appearance of 
an isolated 7-disclination. The white circle is the geodesic circle centered on the 7 that encloses an integrated Gaussian curvature of $-\frac{\pi}{3}$. Right: Nodoid with aspect ratio 
$\rho=0.64$. The geodesic circle, centered on the 7, that encloses $-\frac{\pi}{3}$ integrated Gaussian curvature (white circle) is compared to its lattice analog.}
\end{figure}
 
In Fig.~\ref{cir}(left) we show a minimal energy configuration, with aspect ratio $\rho=0.774$,
containing an isolated 7-fold disclination. We also show the geodesic circle, $\Gamma$,  with radius $r=0.293$, that is the boundary of a disk $D$ whose integrated Gaussian curvature is $-\frac{\pi}{3}$. $\Gamma$ was determined by constructing a geodesic polygon with enough edges that the integrated curvature coincides with the total exterior angle deficit and applying the Gauss-Bonnet theorem
\be \label{GB}\int_DKdA+\sum\limits_i\gamma_i=2\pi,\ee 
where $K$ is the Gaussian curvature and $\gamma_i$ is the exterior angle deficit at vertex $i$. Since disclinations naturally couple to Gaussian curvature in the effective Hamiltonian that controls their energetics \cite{BNT,BG,BY,IVC} a natural criterion for the ground state to admit a defect with a net $-1$ disclination charge is that there be a domain of the surface, centered on the defect, with integrated Gaussian curvature matching the disclination charge $-\pi/3$. 
 
In Fig.~\ref{cir}(right) we show a configuration, with aspect ratio $\rho=0.64$, containing 
a completely isolated 7 along with the geodesic circle centered on the 7 and enclosing an integrated Gaussian curvature of $-\frac{\pi}{3}$.
We have also inscribed a 7-sided geodesic polygon as determined by the triangulation.  
The exterior angles grow from $2\pi/7$ at the 7-fold  disclination to $2\pi/6$ at the inscribed geodesic heptagon. This nodoid has vanishing contact angle at the boundary, 
corresponding to complete wetting of the capillary bridge at the plates; it corresponds to the nodoid with parameter $t$ covering all $\RR$. For smaller aspect ratio the contact angle grows to a maximum of $22^{^{o}}$.  

%To fully understand the relative extension of the solid angle, 
%$\frac{\pi}{3}$, in the geodesic discs, let $N$ be the nodoid with $\rho=0.64$, 
%the quocient between the total curvature of the geodesic disc and the nodoid is 
%$\label{D/N/K}\displaystyle\frac{\int_DKdA}{\int_NKdA}=\frac{1}{12},$
%whereas between the related areas is 
%$\label{D/N/A}\displaystyle\frac{\int_DdA}{\int_NdA}=0.084,$ 
%which represents a ratio area/curvature of 1. 
%The corresponding ratio in the case of the first appearance of an isolated 
%heptagon for $\rho=0.774$ is $1.25$.

\begin{figure}[htb]
\begin{center}
\scalebox{.12}{{\includegraphics{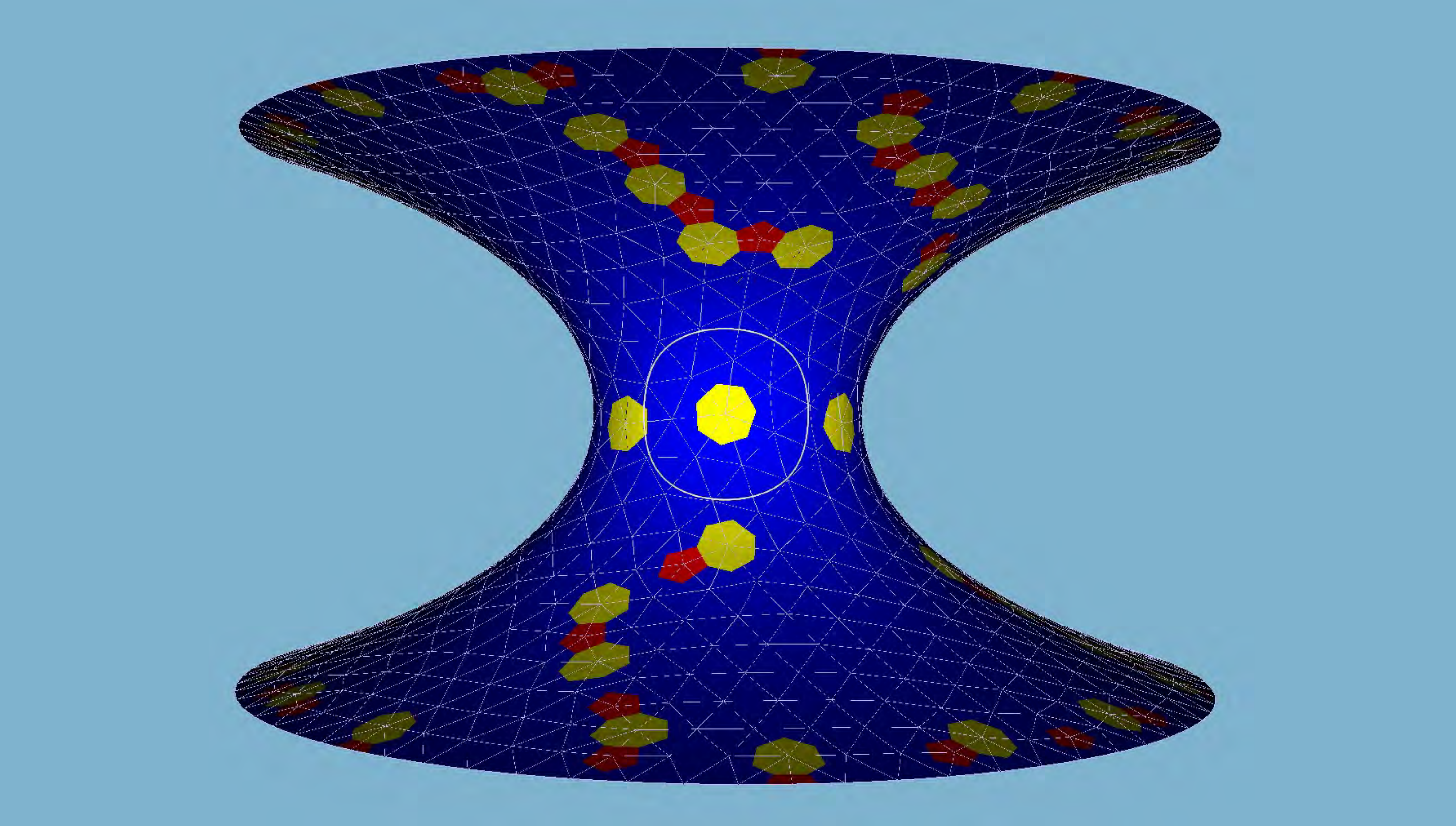}}}
\scalebox{.12}{{\includegraphics{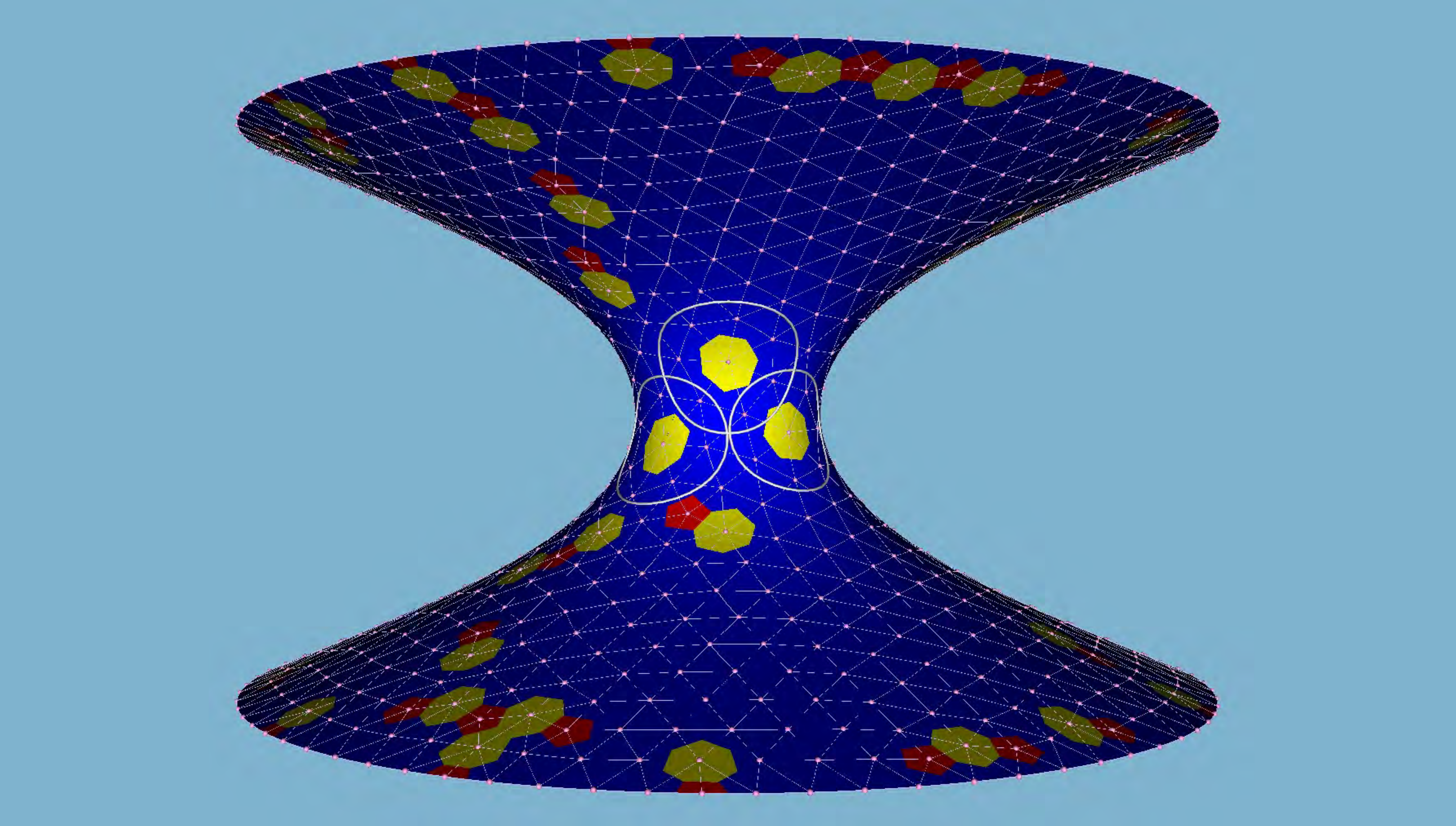}}}
\end{center}
\caption{\label{last}Left: Catenoid (vanishing mean curvature) with aspect ratio $\rho=0.28$. 
Note the appearance of several isolated 7s in the region of the waist. 
The white contour is that of a geodesic circle centered on the 7 inside and enclosing
an integrated Gaussian curvature of $-\frac{\pi}{3}$. Right: Unduloid with aspect ratio $\rho=0.198$. 
Three geodesic disks enclosing an integrated Gaussian curvature of $-\frac{\pi}{3}$ are also shown.}
\end{figure}

As the capillary bridge is stretched to have a very narrow waist (the equatorial section $t=0$ of the Delaunay surface) with large negative maximal Gaussian curvature  we observe a proliferation of 7-fold disclination defects (heptagons). Two examples are displayed in Figs.~\ref{last}(left, right). 
In both configurations there are a total of 11 isolated 7s plus scars, which have the same net disinclination charge. The 7s preferentially occupy the waist of the capillary bridge, whereas scars and pleats are found throughout the surface. 

The sequence of defect motifs revealed by our simulations conforms remarkably closely to that found experimentally in \cite{IVC} (see in particular their Fig.~4). The initial compressed bridge is free of defects. As the bridge is stretched, both experimentally and numerically, one observes the appearance of dislocations, neutral disinclination dipoles, polarized with the 7-fold disclinations towards the maximally negatively curved neck (see Fig.~\ref{95_97} and Fig.4(i) of \cite{IVC}). 
Further stretching leads to the appearance of pleats (see Fig.~\ref{lin81_78} and Fig.4(j) of \cite{IVC}). Finally yet more stretching leads to the appearance of isolated 7-fold disclinations and scars (see Fig.~\ref{cir} and Fig.4(k) of \cite{IVC}).
These comparisons were also noted in \cite{KW} but we note that all the stretched manifolds displaying defects are nodoids (as clearly established in Fig.~\ref{family}) whereas the simulations in \cite{KW} are only for catenoids and unduloids.

\subsection{Stretching from an initial square cylindrical capillary bridge}

In this section we discuss the appearance of defects upon stretching through 
another family of Delaunay surfaces preserving volume and contact radius. 
\begin{figure}[htb]
\begin{center}
\scalebox{.14}{{\includegraphics{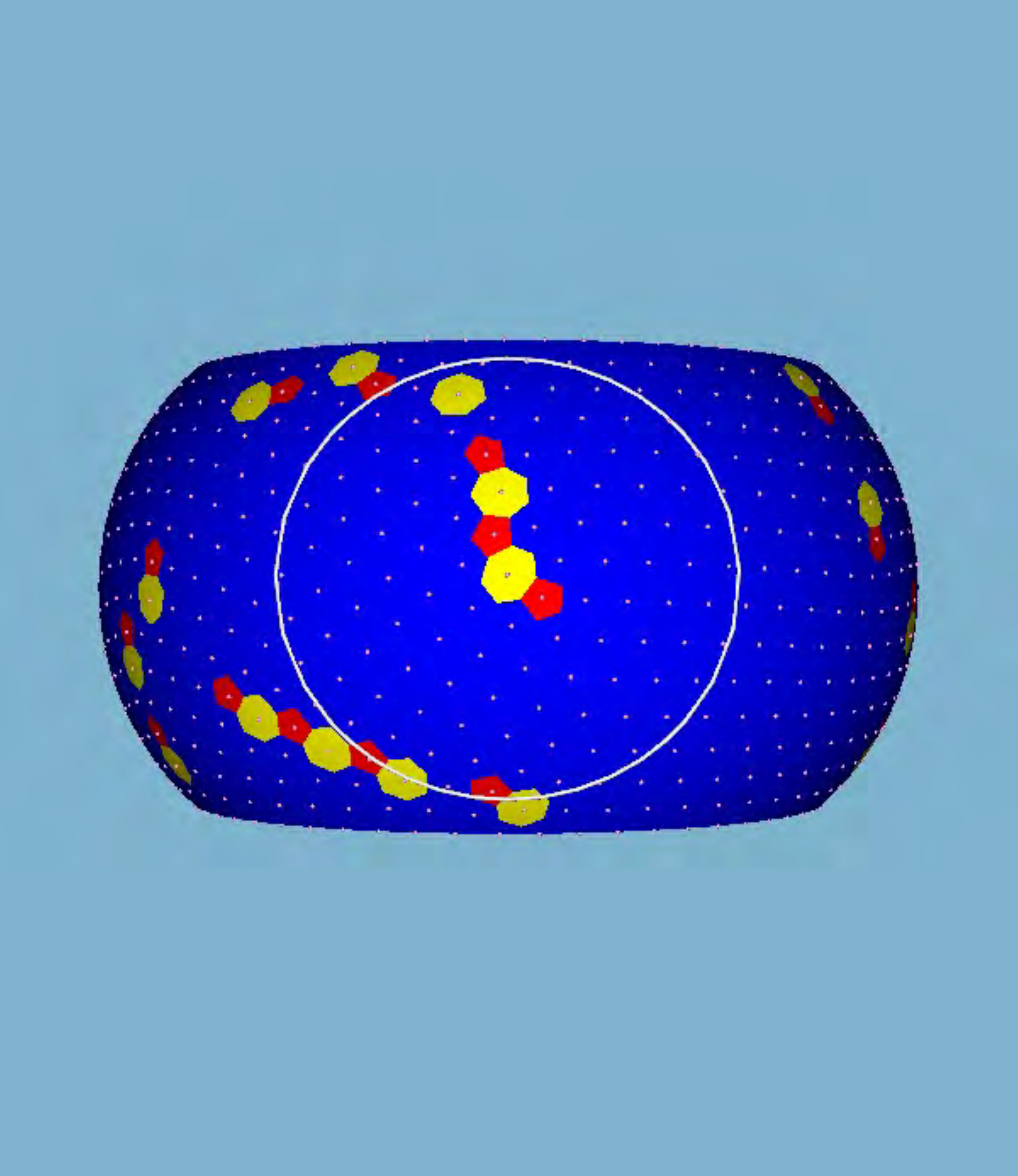}}}
\scalebox{.14}{{\includegraphics{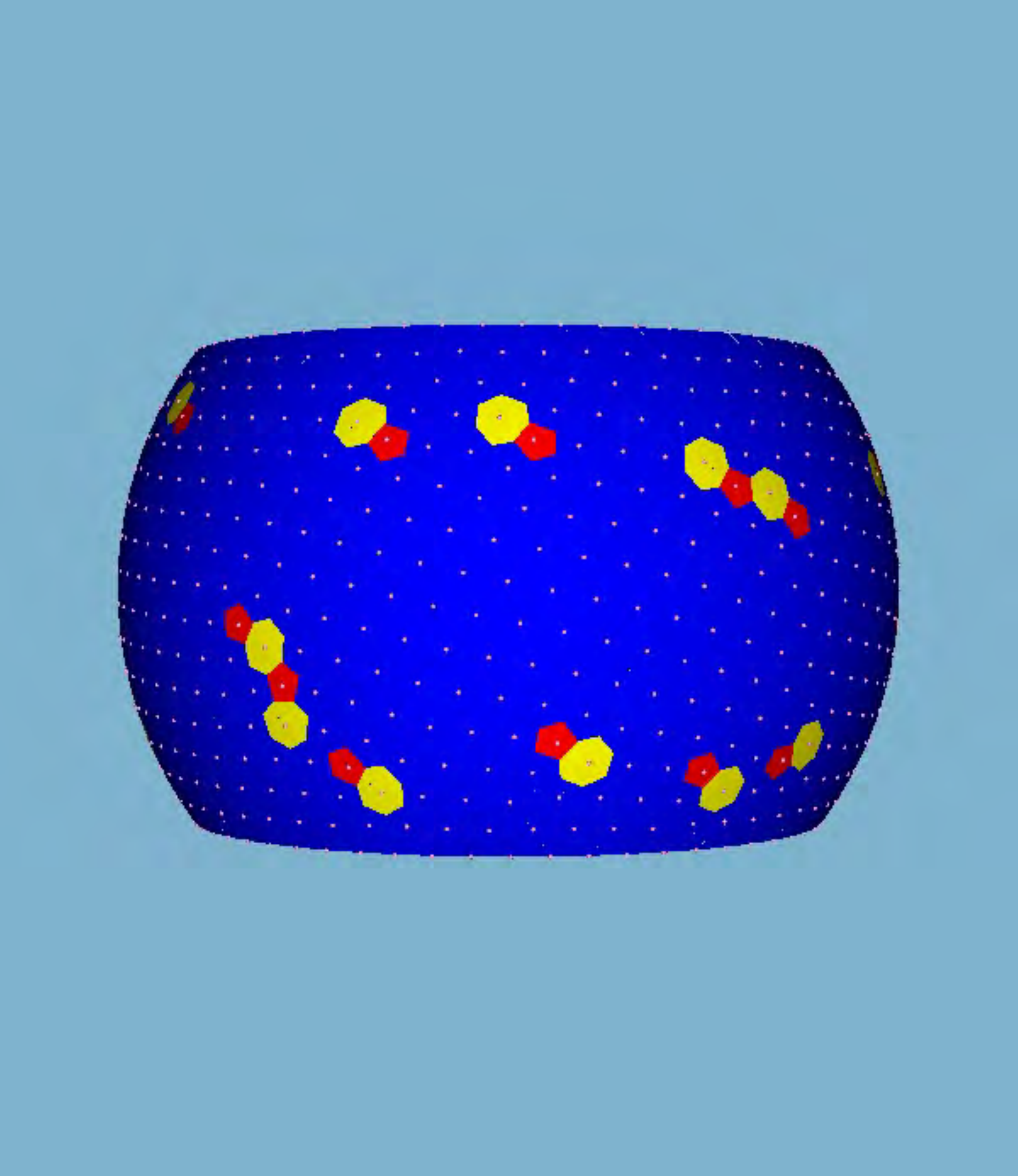}}}
\scalebox{.14}{{\includegraphics{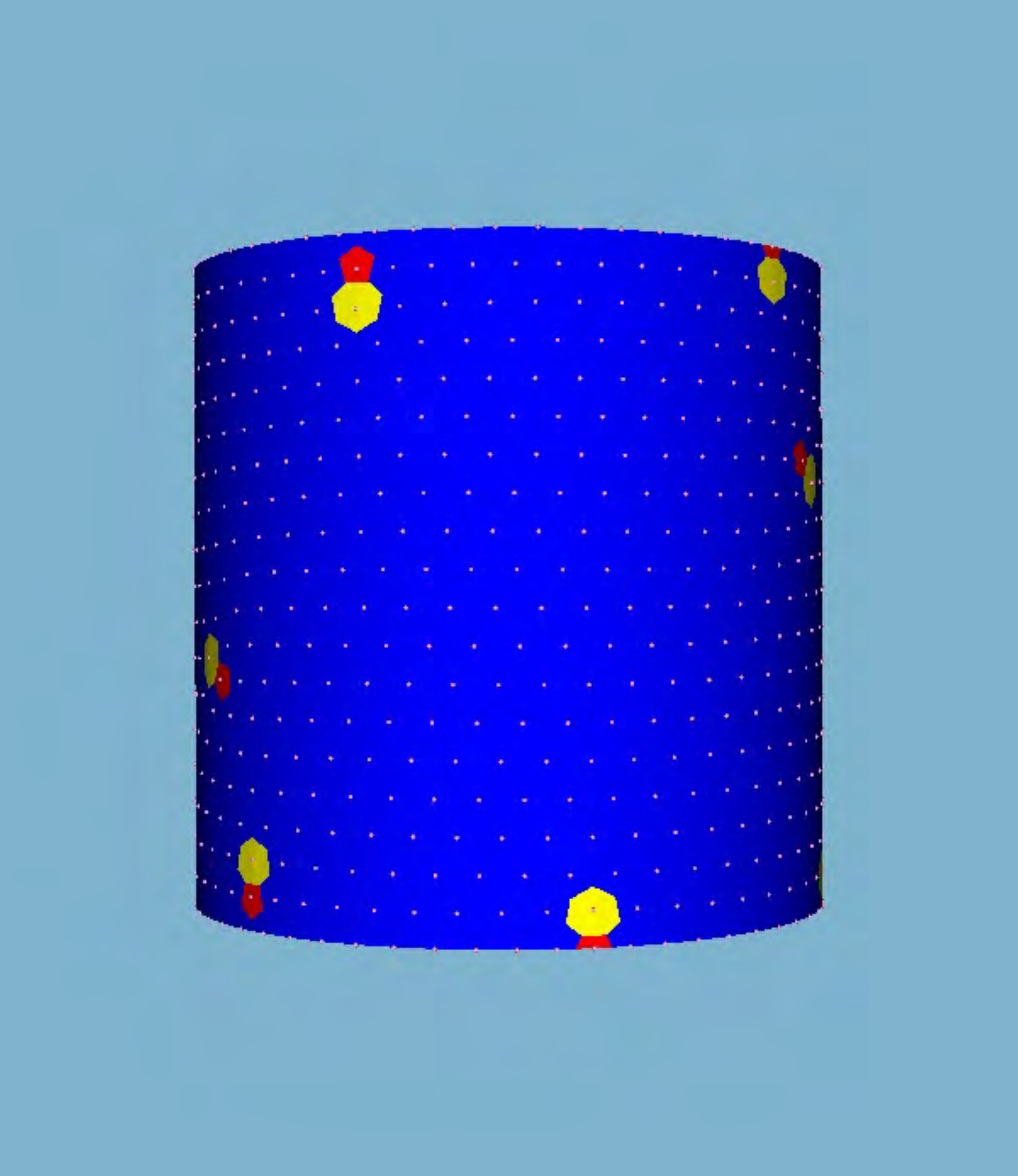}}}
\scalebox{.14}{{\includegraphics{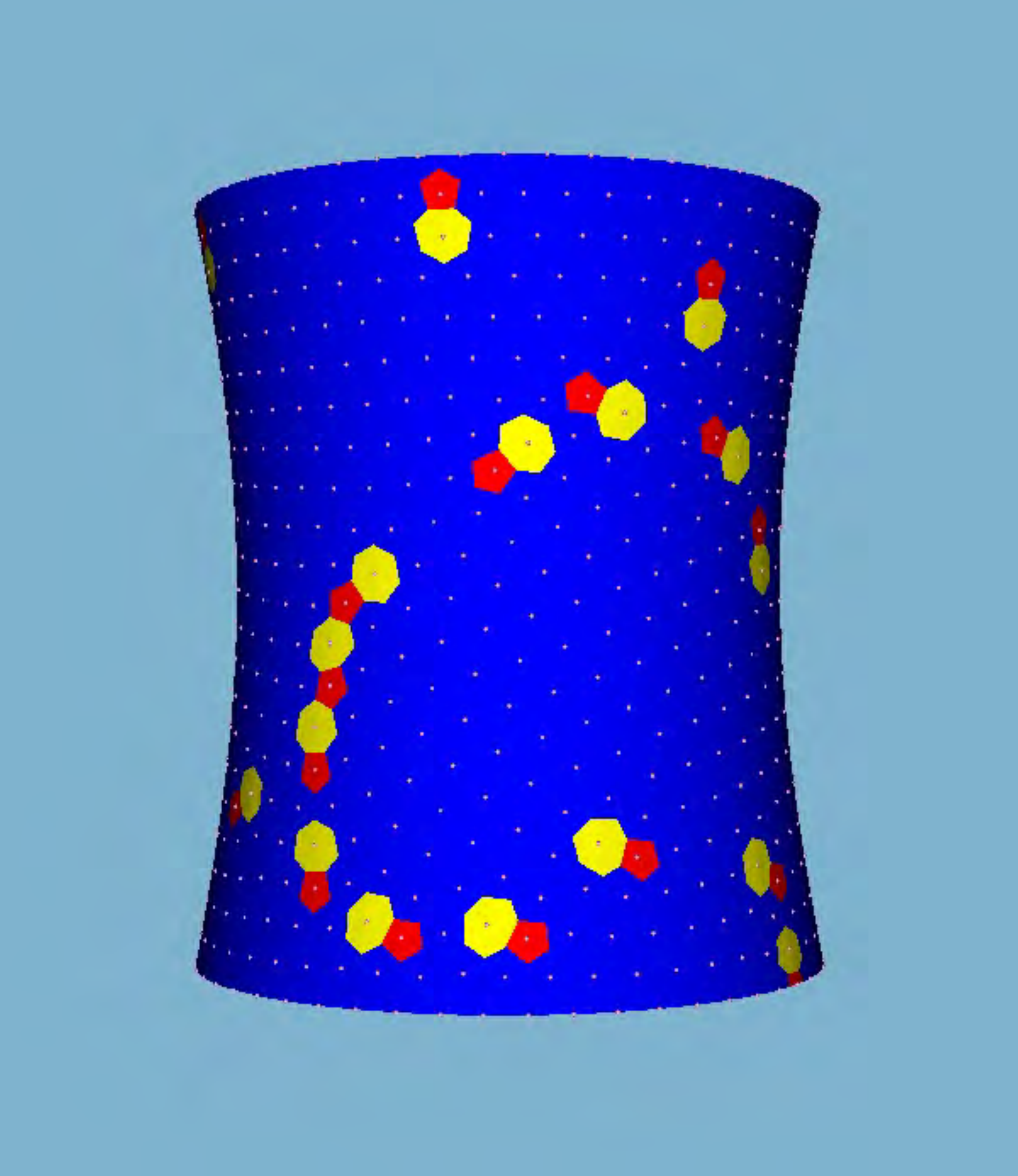}}}
\scalebox{.14}{{\includegraphics{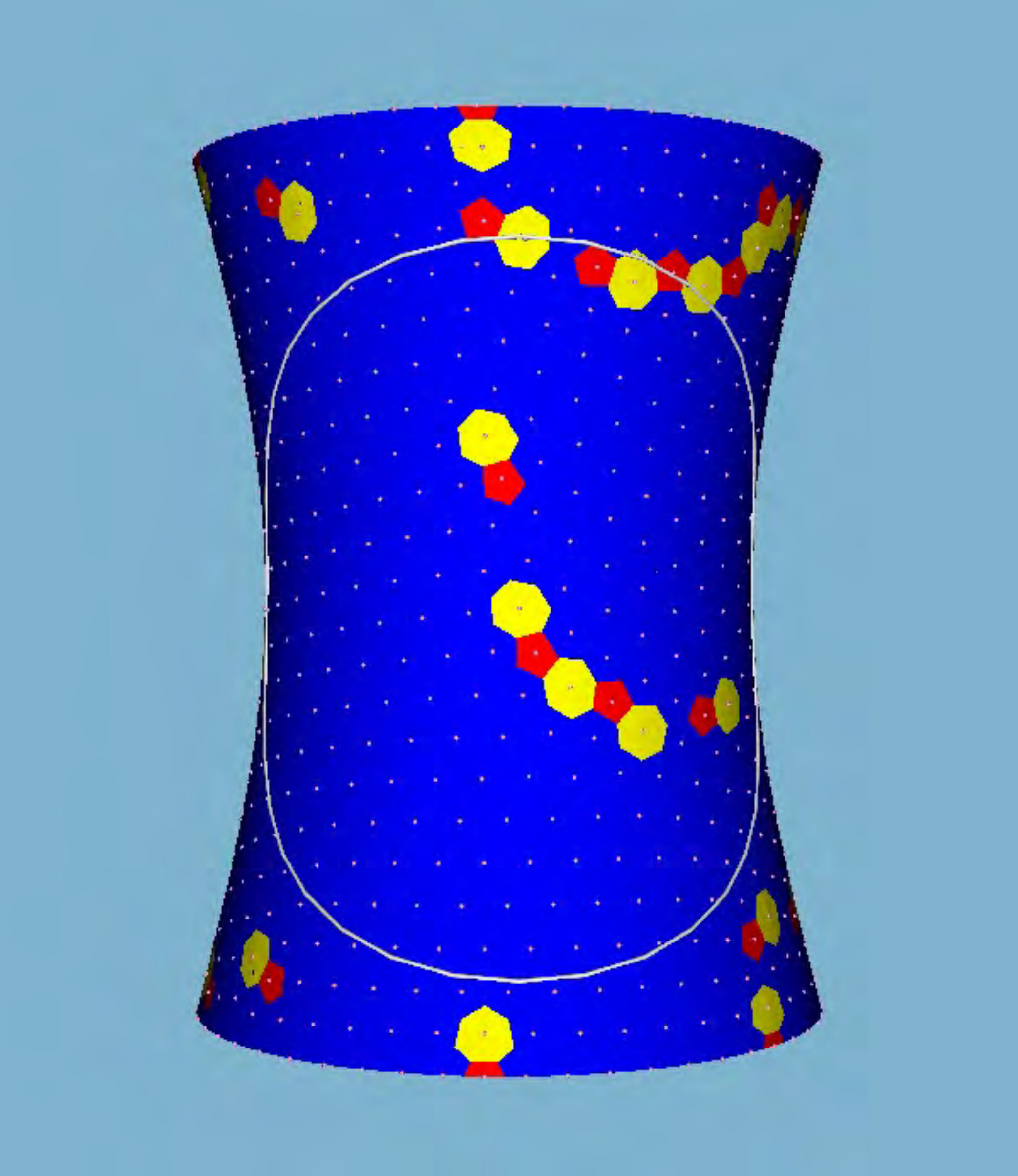}}}
\end{center}
\caption{\label{square} Square cylindrical capillary bridges. (a) Positive Gaussian curvature nodoid with aspect ratio $\rho=1.30$. Observe the encircled 5-7-5-7-5 scar separated by a lattice spacing from an isolated 7. The geodesic disc enclosed by the circle has integrated Gaussian curvature $\frac{\pi}{3}$.
(b) Positive Gaussian curvature unduloid with aspect ratio $\rho=1.24$. Observe isolated dislocations and pleats. (c) Null Gaussian curvature cylinder with aspect ratio $\rho=1$. (d) Negative Gaussian curvature nodoid with aspect ratio $\rho=0.86$. Note the long 7-5-7-5-7-5 pleat situated left center. (e) Negative Gaussian curvature unduloid with aspect ratio $\rho=0.77$. Observe the prominent 7-5-7-5-7 scar. The geodesic disc enclosed by the circle has integrated Gaussian curvature $-\frac{\pi}{3}$.}
\end{figure}
This family is generated by starting from a positive Gaussian curvature nodoid with 
aspect ratio $\rho=1.30$ and is designed to pass through the square cylindrical capillary 
bridge with $r_c=1$ and $h=2$ (aspect ratio $\rho=1$). It terminates in an unduloid with 
aspect ratio $\rho=0.77$. 
Representative configurations for this sequence are shown in Fig.~\ref{square}. 
The central cylinder exhibits very few defects. Increasing or decreasing the aspect ratio leads 
first to the appearance of pleats and then to longer pleats and scars.  Geodesic discs with total 
curvature $\pm\frac{\pi}{3}$ are shown surrounding scars and are completely
contained in the interior of the surface. Note that the defect structure in Fig.~\ref{square}(a) is a pleat with the top 7-fold disinclination separated by one lattice spacing from the rest of the pleat.

\section{Minimum energy configurations on slices}

\begin{figure}[htb]
\begin{center}
\scalebox{.17}{{\includegraphics{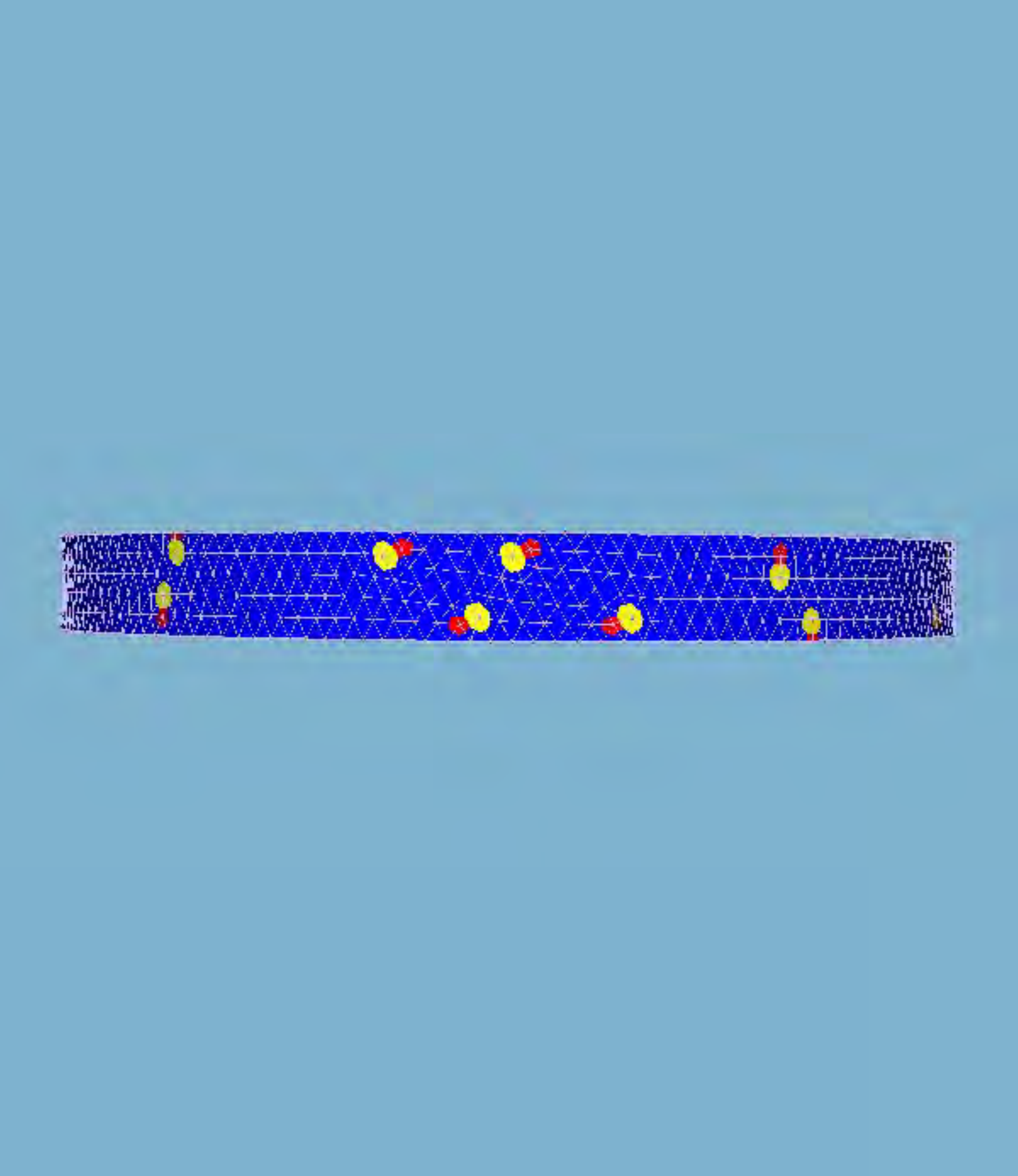}}}
\scalebox{.17}{{\includegraphics{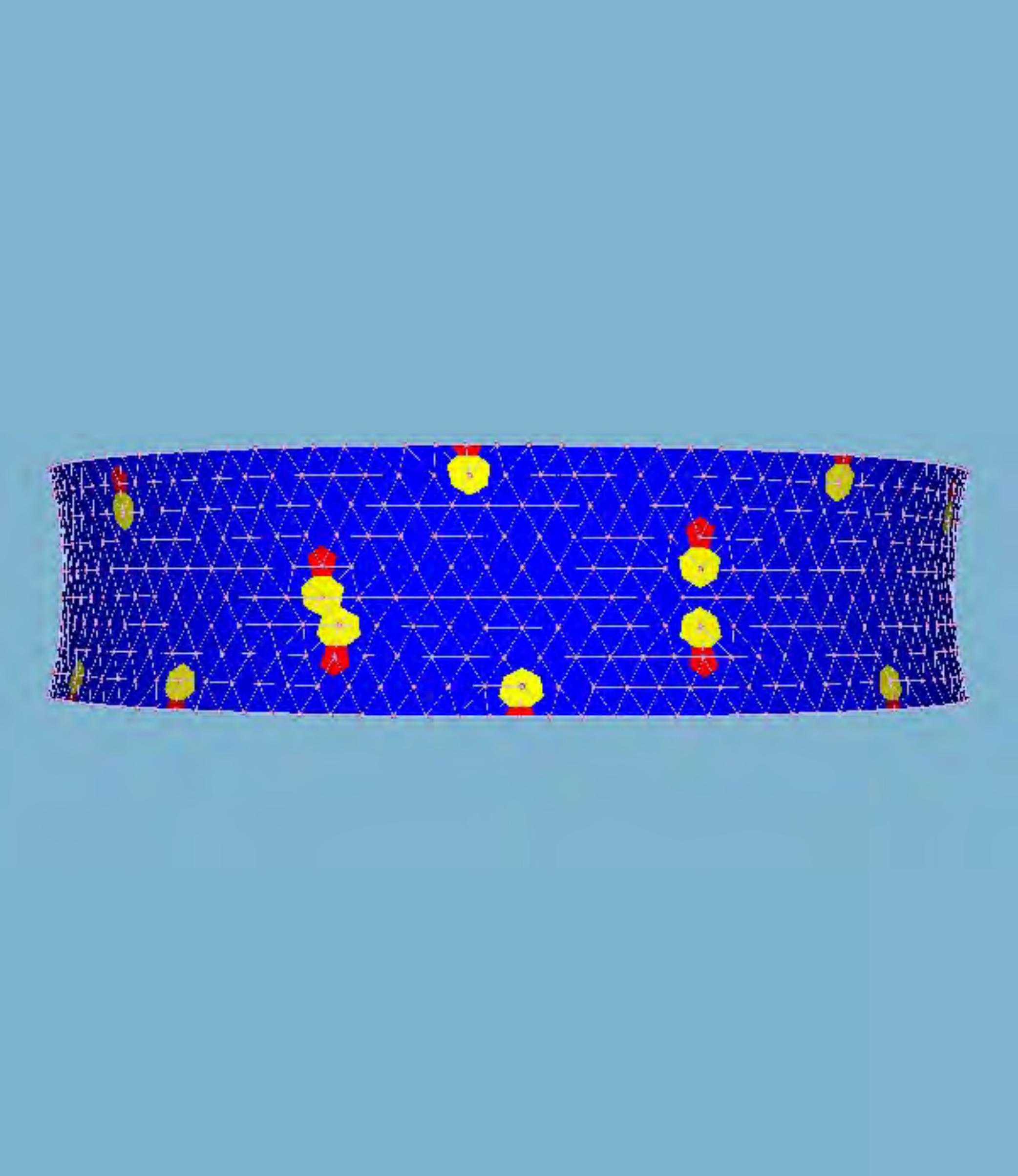}}}
\scalebox{.17}{{\includegraphics{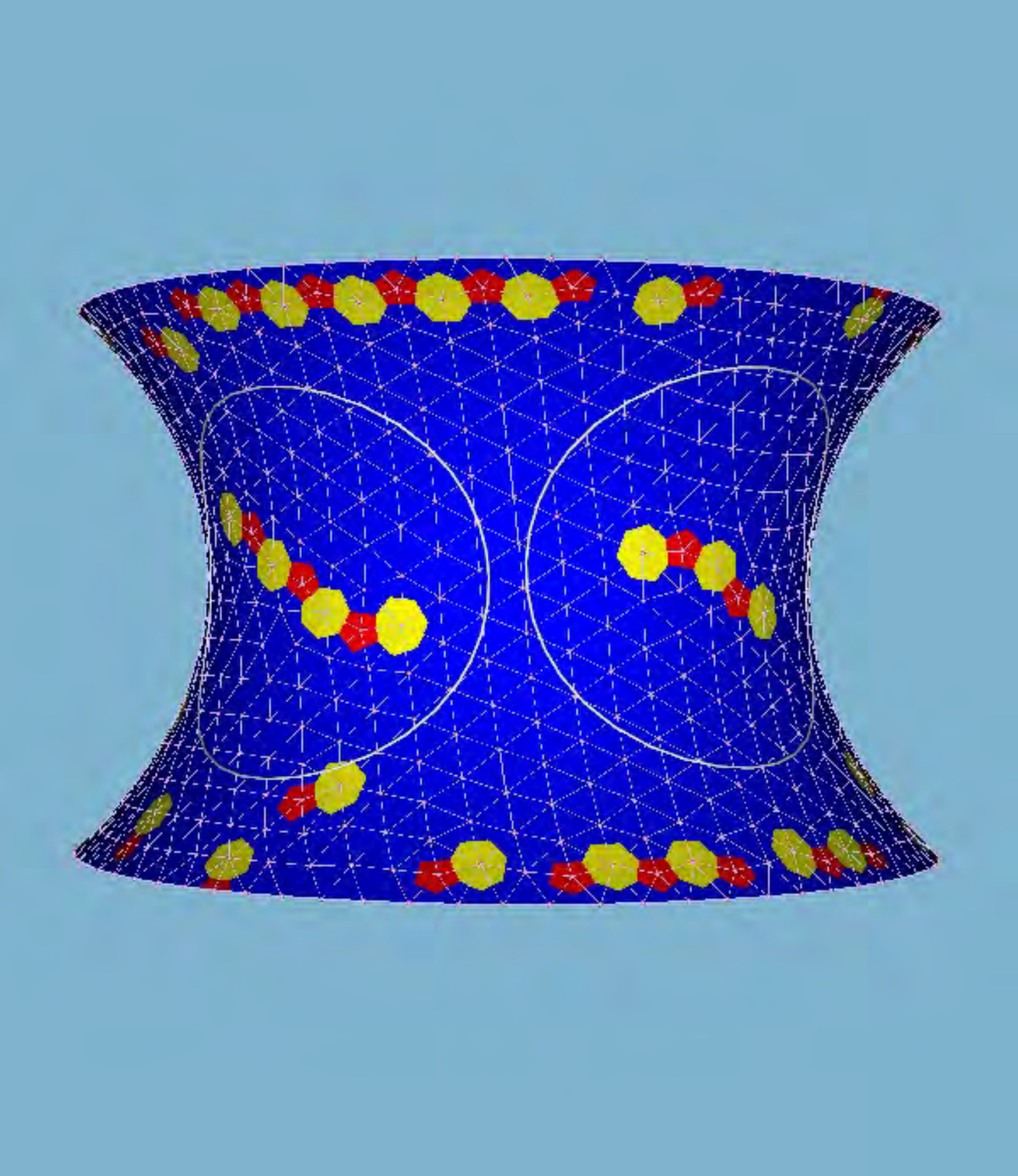}}}
\scalebox{.17}{{\includegraphics{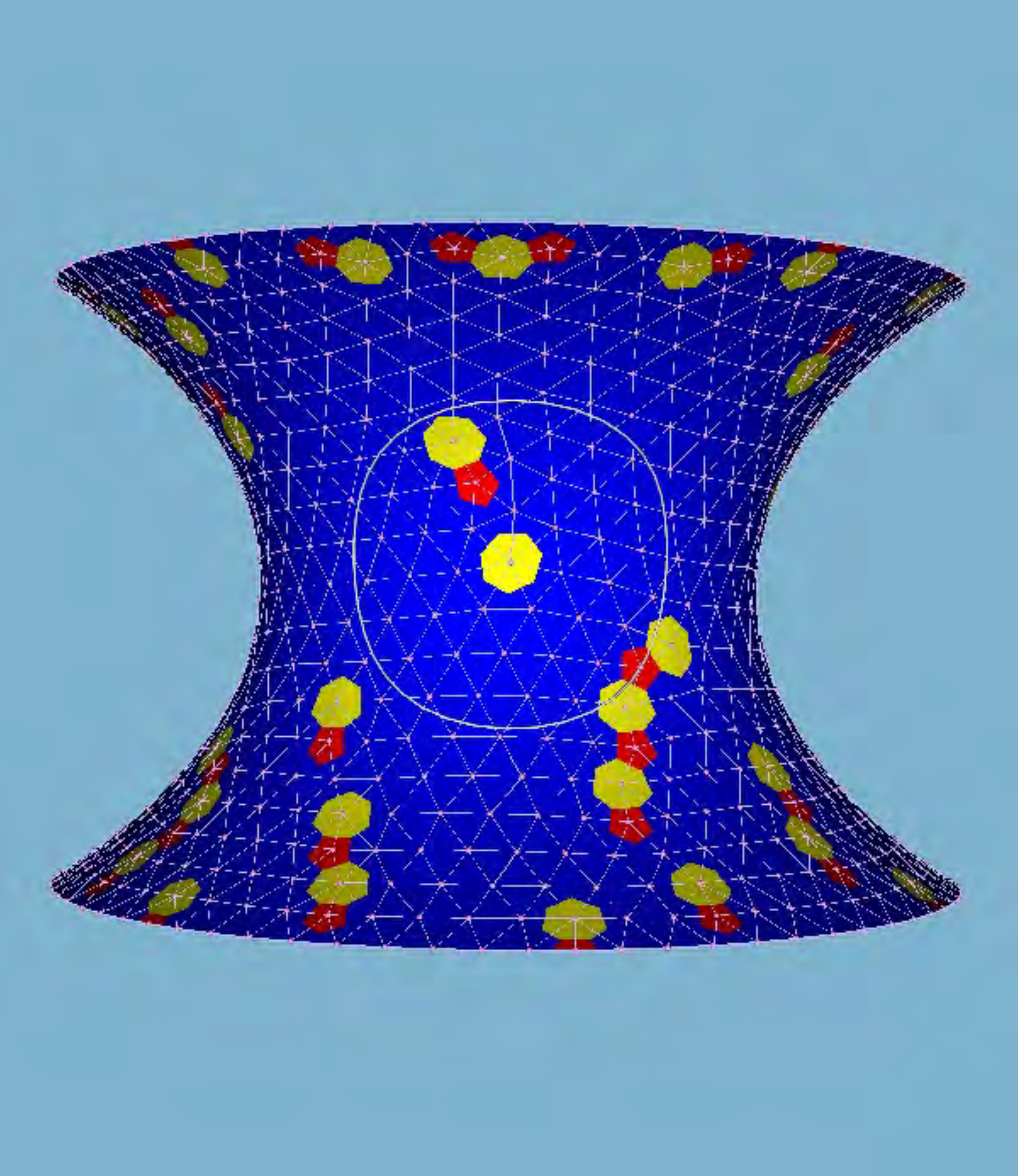}}}
\end{center}
\caption{\label{0501} Different slices of the nodoid with parameters $a=0.5$ and $b=0.1$. (a) $t_{max}=0.02$, (b) $t_{max}=0.05$, (c) $t_{max}=0.2$, (d) $t_{max}=0.3$.}
\end{figure}

For an isolated 7-fold disclination (heptagon) to be present in minimum energy configurations of crystalline arrays on a negative curvature surface a reasonable criterion is that the integrated Gaussian curvature over the geodesic disc centered at the disclination be more negative than $-\frac{\pi}{3}$, the deficit angle corresponding to the topological charge of a 7-fold disclination~\cite{BG}.
To check that it is the integral of the Gaussian curvature, instead of the
Gaussian curvature itself, that determines the character of defect motifs we
examine minimum energy configurations on slices of a given Delaunay surface with steadily increasing values of $t_{max}$, the maximal value of the meridian coordinate $t$. Each slice has $t$ spanning the interval $[-t_{max},t_{max}]$. 
The shape of a nodoid is completely determined by the parameters $a$ and $b$. 
We merely restrict to a growing set of slices by varying $t_{max}$. 
The first surface we investigate is the nodoid with $a=0.5$ and $ b=0.1$.  
This capillary bridge has maximal Gaussian curvature at the $t=0$ waist (equator) of $K(0)=-10,401$. 
In Fig.~\ref{0501}(a) we take a slice with $t_{max}=0.02$ and the behavior is completely equivalent to a cylinder. For completeness  we note that this surface encloses a volume $V=6.11\cdotp 10^{-7}$, 
has a lateral area $A=1.23\cdotp 10^{-4}$ and an equilibrium potential energy 
 $E=1.17\cdotp 10^{13}$.  

In Fig.~\ref {0501}(b) we show a slice with $t_{max}=0.05$. 
For this slice, dislocations are found in the interior just as in the case of stretching 
within the fat cylinder family. The evolution of defect motifs continues with the 
appearance of pleats and, in Fig.~\ref {0501}(c), where $t_{max}=0.2$, 
disclination charge $-1$ scars sitting inside discs of integrated Gaussian curvature 
$-\frac{\pi}{3}$. The first appearance of an isolated 7-fold disclination (heptagon) is for a slice  
with $t_{max}=0.3$. In Fig.~\ref {0501}(d) we show the geodesic disc 
of geodesic radius $r=0.0061$, the same value as in the previous 
slice, because the surfaces are the same except for their height. 
This last surface encloses a volume of $V=1.17\cdotp 10^{-5}$ and has a 
lateral area  $A=3.39\cdotp 10^{-3}$.

As we increase the parameter $b$, the  Gaussian curvature in the waist
is strongly reduced and yet the sequence of defect motifs is essentially the same.  
To be concrete we show the family of slices with $a=0.5$ and $b=0.4$. 
Now the Gaussian curvature in the waist is $K(0)=-65$.
\begin{figure}[htb]
\begin{center}
\scalebox{.17}{{\includegraphics{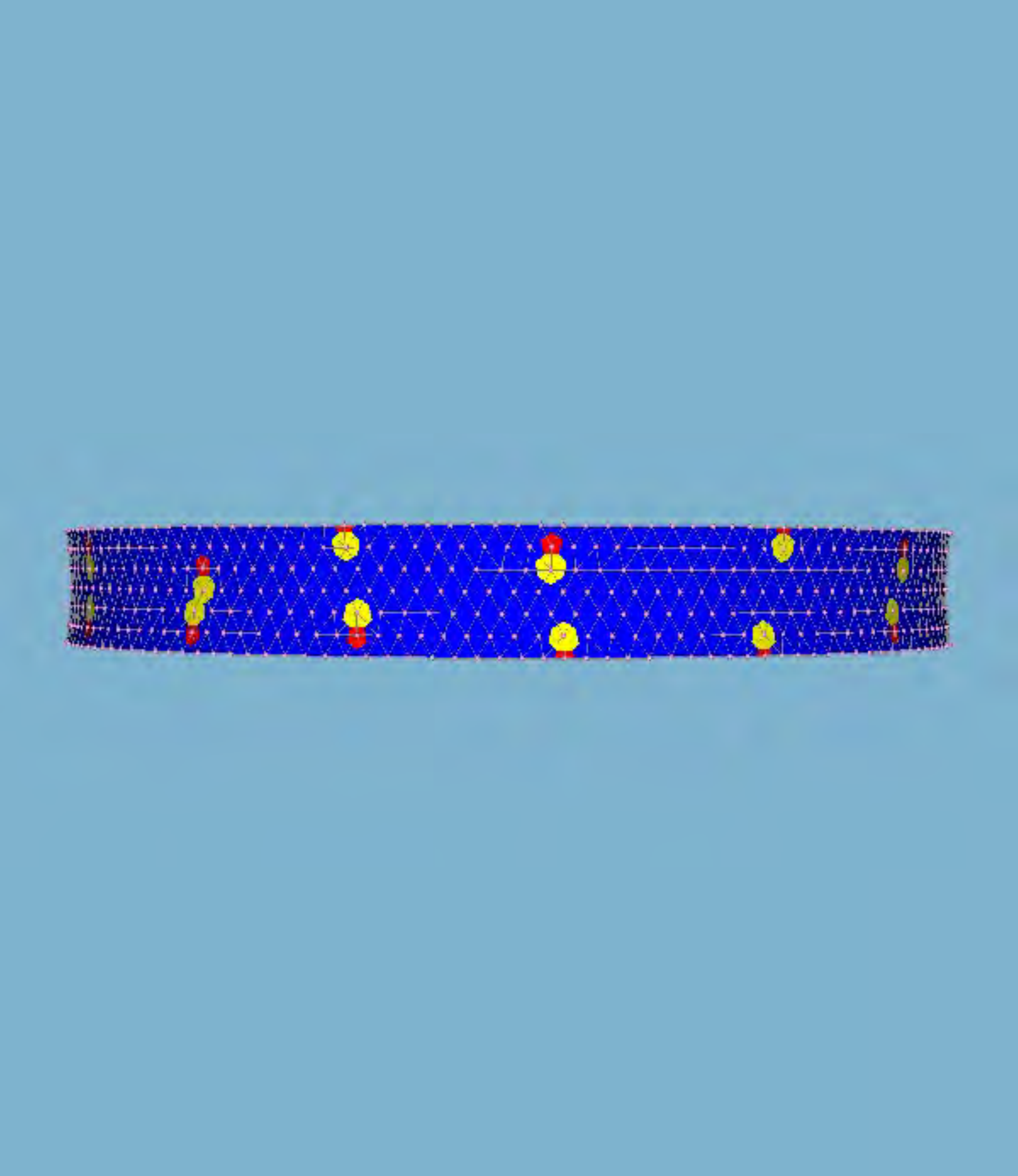}}}
\scalebox{.17}{{\includegraphics{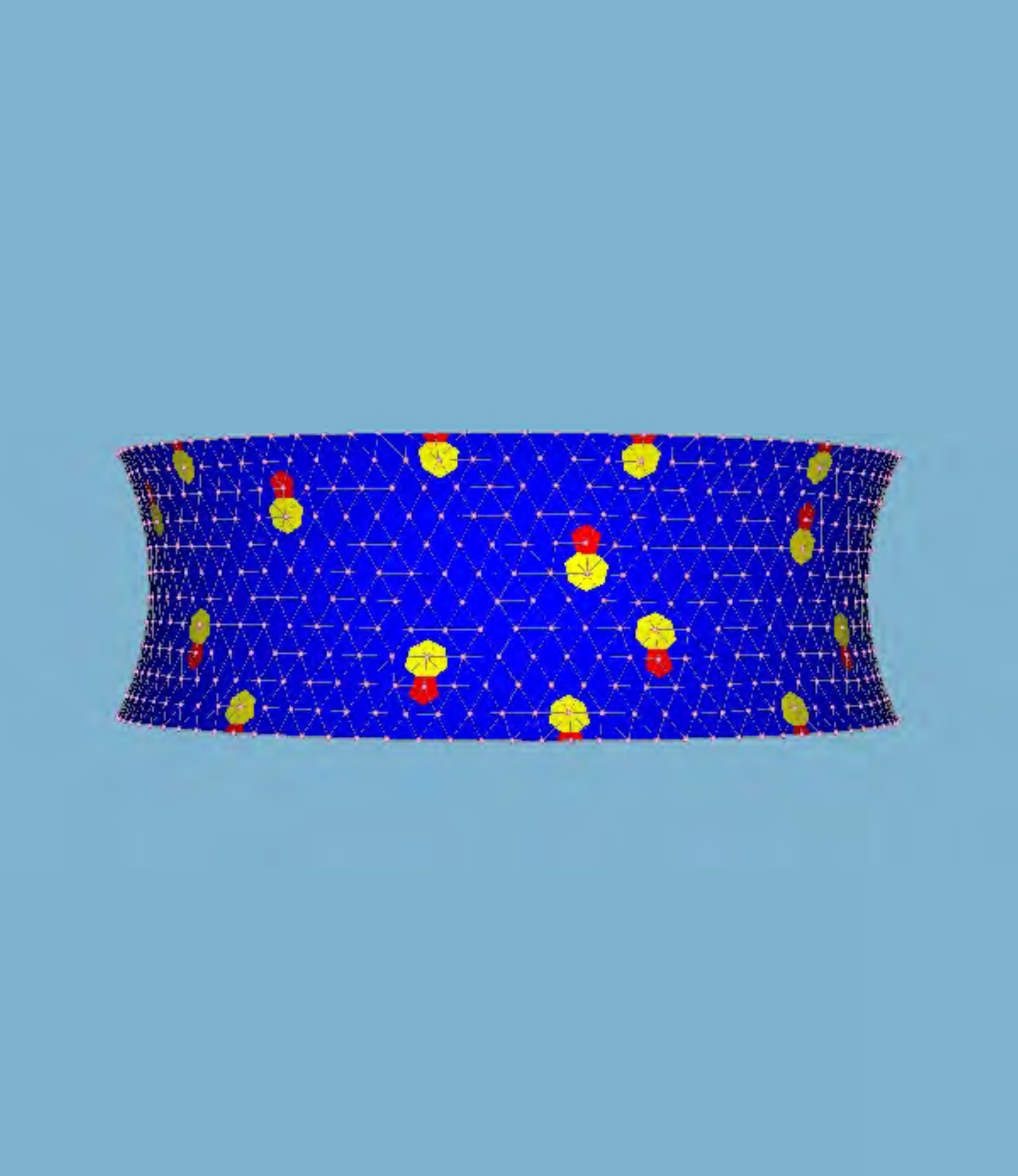}}}
\scalebox{.17}{{\includegraphics{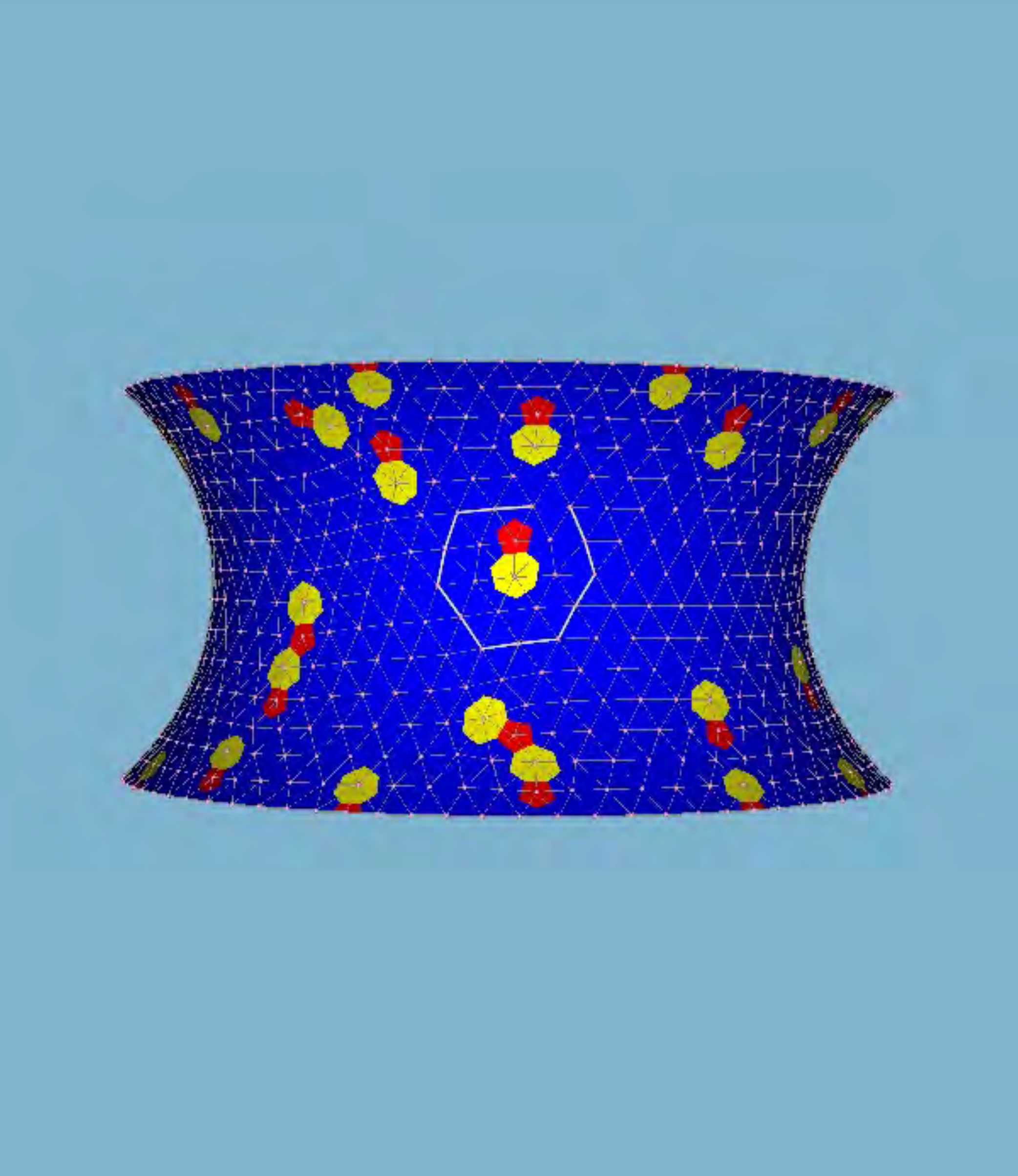}}}
\scalebox{.17}{{\includegraphics{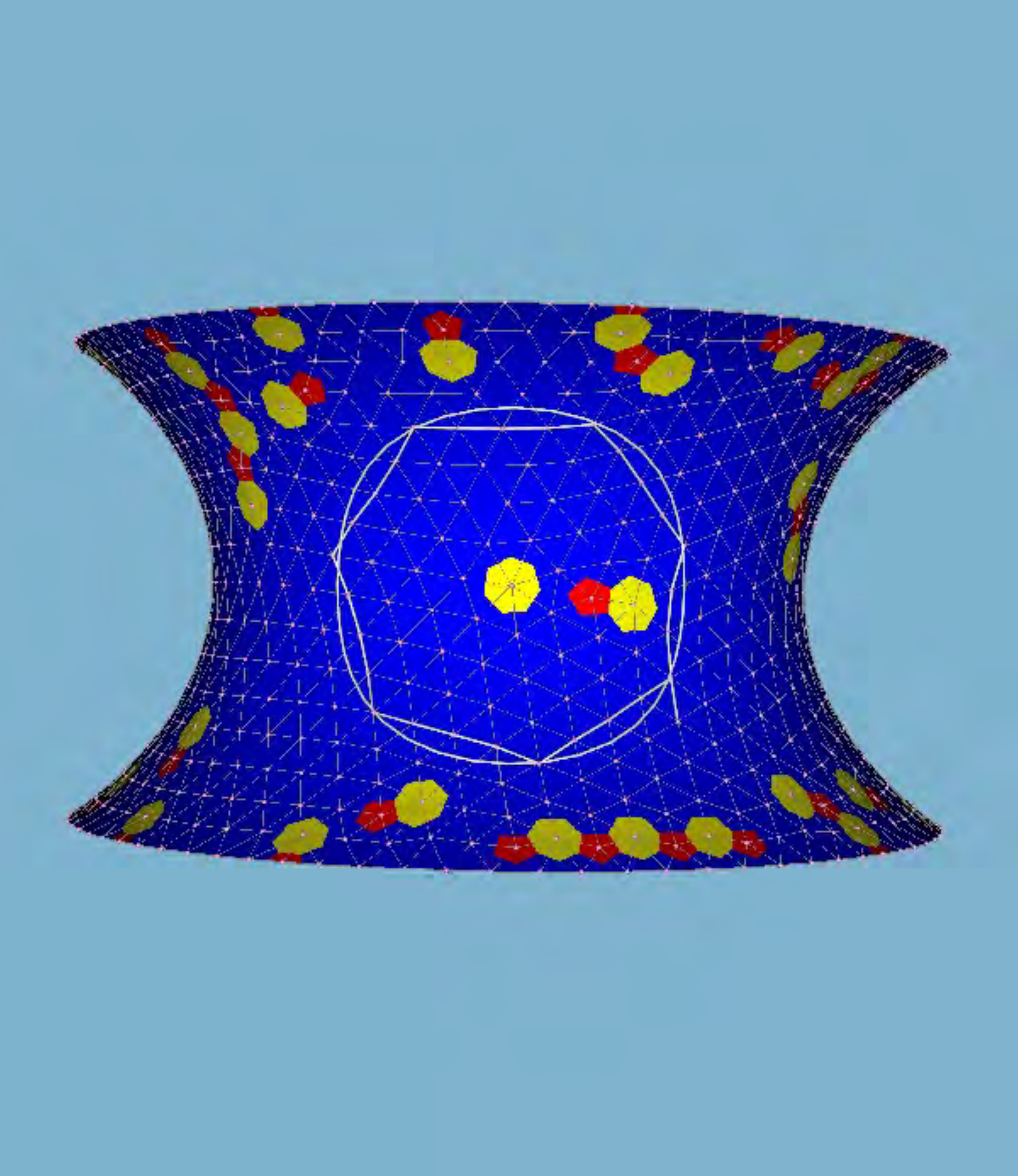}}}
\end{center}
\caption{\label{0504}Different slices of the nodoid with $a=0.5$ and $b=0.4$. (a) $t_{max}=0.1$, (b) $t_{max}=0.3$, (c) $t_{max}=0.6$, (d) $t_{max}=0.8$.}
\end{figure}

In Fig.~\ref {0504}(a) we display a slice with $t_{max}=0.1$ and the 
pattern is again equivalent to a cylinder. Its volume is 
$2.17\cdotp 10^{-3}$ and the lateral area  is $3.1\cdotp 10^{-2}$.
In Fig.~\ref{0504}(b) we again observe the formation
of dislocations in the interior. In Fig.~\ref{0504}(c) we have 
enclosed a dislocation with a polygon to show the associated Burgers' vector. 
In Fig.~\ref{0504}(d) we see an isolated 7-fold disclination with a nearby dislocation.

%On the other hand, in the case $a=0.5,\, b=0.1,\, t_{max}=0.3$, the geodesic 
%radius  of the circle around the isolated heptagon is $r=0.0061$, whereas 
%the average side size of the geodesic triangulation  is $l_m=0.00171$ and 
%we have as  ratio $\frac{r}{l_m}=3.57$. For the case 
%$a=0.5,\, b=0.4,\, t_{max}=0.8$, 
%in which there is also the first appearance of an isolated heptagon, 
%$r=0.077$ while $l_m=0.01853$, so $\frac{r}{l_m}=4.16$. 
%Observe that in the case of constant volume, 
%the first appearance of an isolated heptagon in $\rho=0.774$ was with 
%ratio $\frac{r}{l_m}=4.17$.
\begin{figure}[htb]
\begin{center}
\scalebox{.12}{{\includegraphics{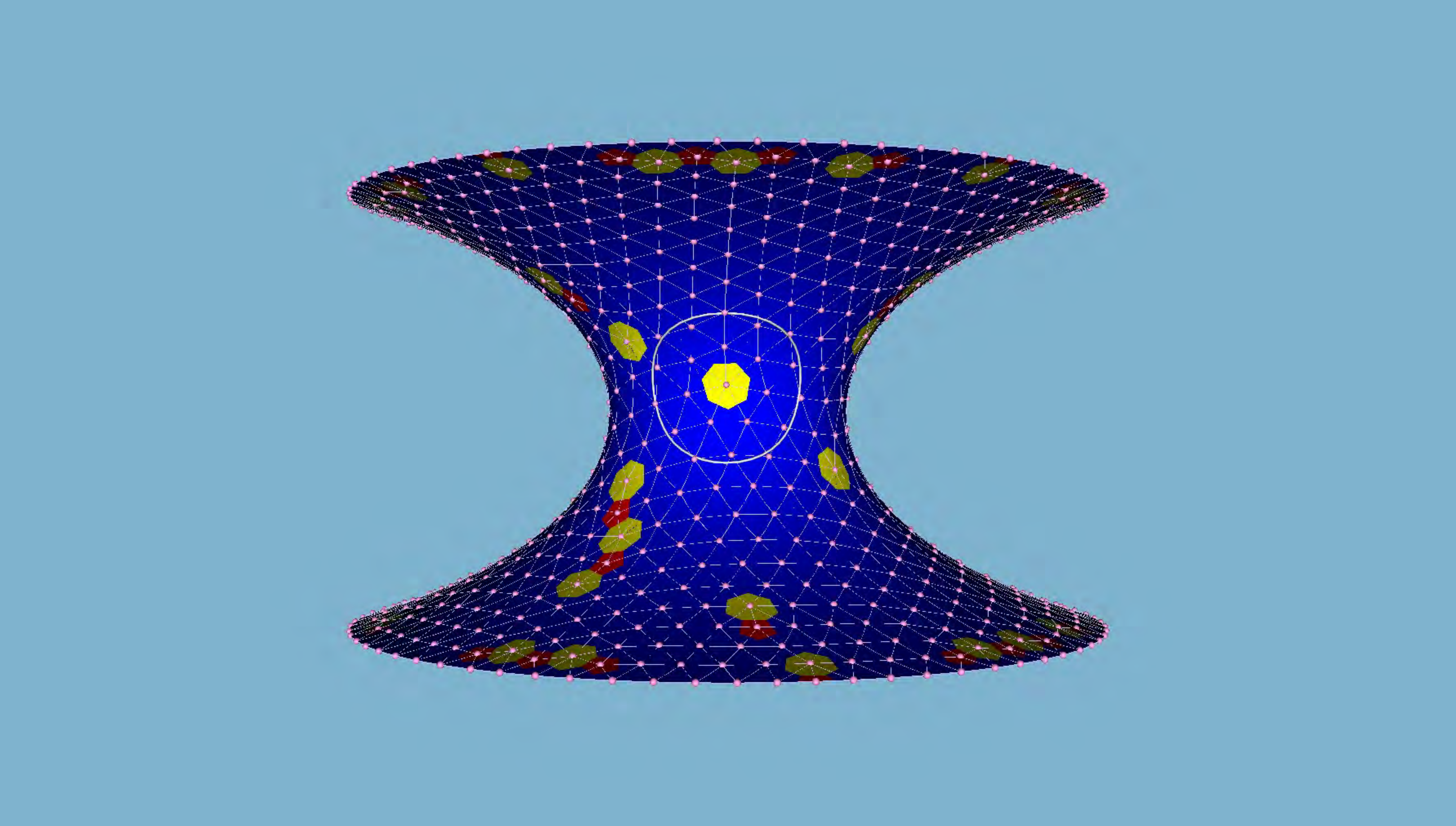}}}
\scalebox{.12}{{\includegraphics{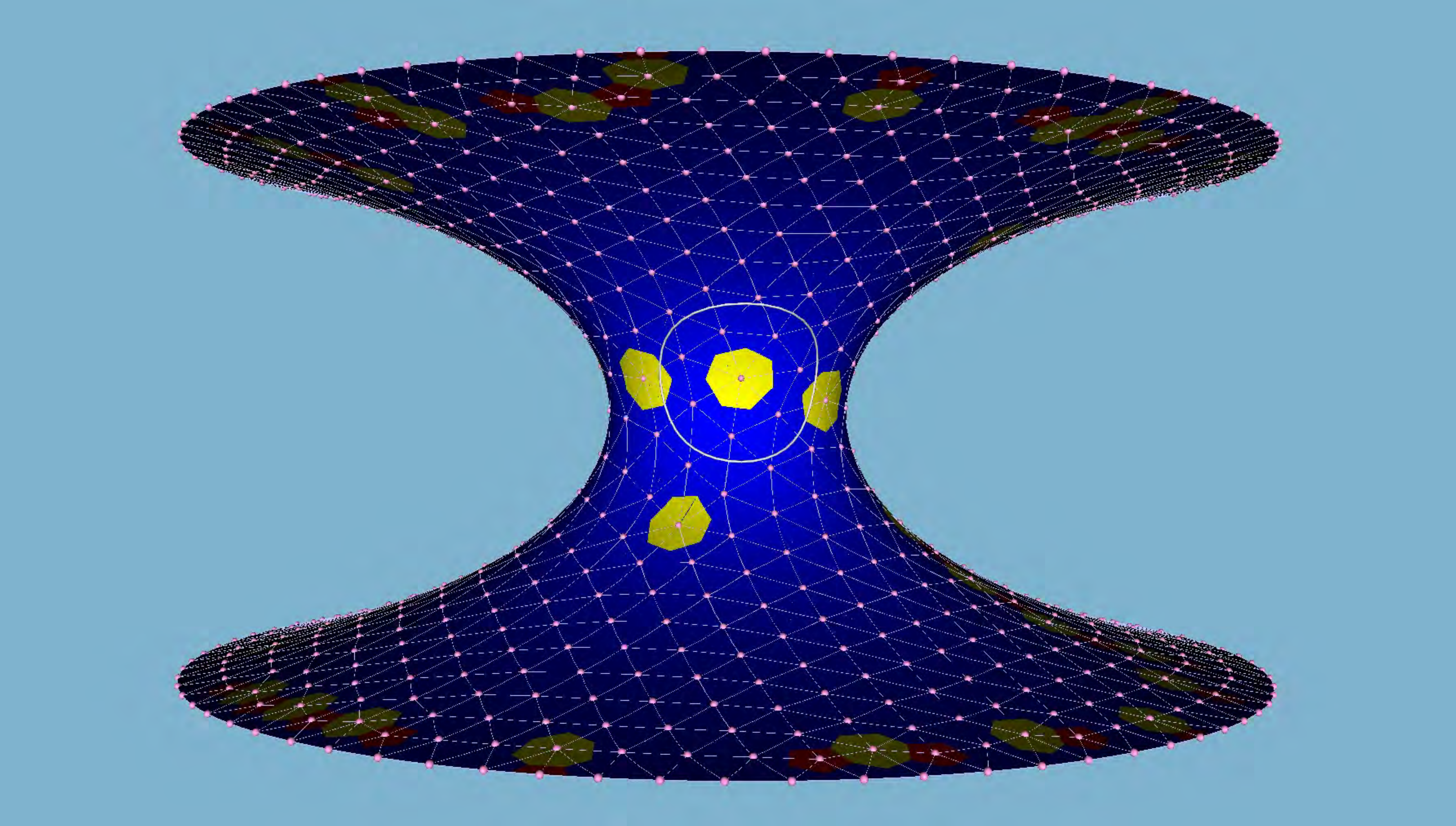}}}
\end{center}
\caption{\label{lastslices} Left: Nodoid with  $a=0.5$, $b=0.1$ and $t_{max}=0.6$ 
with a geodesic disc  of total curvature  $-\frac{\pi}{3}$ in one of the 
isolated 7s in the waist. Right: The same nodoid but with a larger slice, 
now $t_{max}=0.9$.}
\end{figure}
Finally in Fig.~\ref{lastslices} we include the evolution of defects for 
the nodoid with parameters $a=0.5$ and $b=0.1$, when $t_{max}$ increases and the 
slices contain most of the nodoid. In both cases there are  11 total defects 
counting isolated 7s and scars.  In Fig.~\ref{lastslices} (right) the isolated 7s are more concentrated in the waist. 

This we find the full range of phenomena explored experimentally in \cite{IVC} by simulating the full range 
of Delaunay surfaces generated in stretching. We emphasize that catenoids and unduloids alone are insufficent in making a proper comparison with experiment \cite{KW} since the majority of surfaces encountered in stretching are nodoids with negative mean curvature as opposed to the positive mean curvature of unduloids.   

The presence of isolated 7-fold disclinations (or equivalently their dual
heptagons) or scars in the ground state offers new possibilities for
supramolecular chemistry via functionalization of crystalline arrays on these
surfaces, provided one can chemically detect the 7-fold disclination (or scars)
and attach ligands there just as smectic disclinations are functionalized by
place-exchange reactions for nanoparticles coated with an equal mixture of short
and long chain alkanethiols in the work of DeVries et al. \cite{DVetal}.
Isolated 7-fold discinations, scars and pleats may also be sources of material
weakness or perhaps improved performance through capturing material dislocations
and thus limiting plastic deformation.

\subsection*{Acknowledgements}

This research work was partly supported by the Spanish Research Council under project MTM2010-19660, by the National Science Foundation through award DMR-0808812 and by the Soft Matter Program at Syracuse University. MJB would like to thank David Nelson and William Irvine for many stimulating discussions on the nature of condensed matter order on capillary bridges over a period of several years.

%\bibliography{CMC_ref}
%\bibliography{CMC_ref}

\end{document}